\documentclass[prd,amsmath,nobibnotes,nofootinbib,superscriptaddress,preprintnumbers]{revtex4}

\usepackage{bm} \usepackage{graphicx}
\usepackage{epstopdf}

\usepackage{epsfig}
\def\ba{\begin{eqnarray}}
\def\ea{\end{eqnarray}}
\def\be{\begin{equation}}
\def\ee{\end{equation}}
\def\({\left(}
\def\){\right)}
\def\[{\left[}
\def\]{\right]}
\def\<{\left<}
\def\>{\right>}
\def\calh{\mathcal{H}}
\def\tr{\mathop{\rm Tr}\nolimits}
\newcommand{\rar}{\rightarrow}

\begin{document}

\title{Out of equilibrium: understanding cosmological evolution to lower-entropy states}
\date{\today}

\author{Anthony Aguirre}
\email{aguirre@scipp.ucsc.edu}
\affiliation{SCIPP, University of California, Santa Cruz, CA 95064, USA}
\author{Sean M. Carroll}
\email{seancarroll@gmail.com}
\affiliation{California Institute of Technology, Pasadena, CA 91125, USA}
\author{Matthew C. Johnson}
\email{mjohnson@perimeterinstitute.ca}
\affiliation{Perimeter Institute for Theoretical Physics, Waterloo, Ontario N2L 2Y5, Canada} 

\begin{abstract}
Despite the importance of the Second Law of Thermodynamics, it is not absolute. Statistical
mechanics implies that, given sufficient time, systems near equilibrium will spontaneously
fluctuate into lower-entropy states, locally reversing the thermodynamic arrow of time.  We study the time
development of such fluctuations, especially the very large fluctuations relevant to cosmology.
Under fairly general assumptions, the most likely history of a fluctuation out of equilibrium is 
simply the CPT conjugate of the most likely way a system relaxes back to equilibrium. 
We use this idea to elucidate the spacetime structure of various fluctuations in (stable and 
metastable) de Sitter space and thermal anti-de Sitter space.
\end{abstract}

\preprint{CALT 68-2845}

\maketitle

\section{Introduction}

The law of non-decreasing entropy is one of the most fundamental in physics.  As Eddington proscribed: ``...if your theory is found to be against the Second Law of thermodynamics I can give you no hope; there is nothing for it but to collapse in deepest humiliation."  Yet in deriving thermodynamics as a limit of statistical mechanics, it becomes clear that the Second Law is not quite inviolable: in an equilibrium system downward excursions in entropy do occur, have been observed in the laboratory~\cite{Wang:2002p3619}, and may indeed be important for the  functioning of life at the molecular level~\cite{PhysRevLett.104.218103}.  In statistical mechanics, the Fluctuation Theorem
quantifies the relative probability that an isolated system will evolve upward or downward in entropy \cite{Evans:1993p3620,doi:10.1080/00018730210155133}. 
Because an excursion decreasing entropy by $\Delta S$ is suppressed by a factor of $\exp(-\Delta S)$,
in macroscopic systems with thousands of particles or more, significant ``miraculous" downward entropy excursions are so rare that they are essentially never important.

The evolution of our observable universe appears to be a steady march towards successively \emph{higher}-entropy states. If the observed dark energy is a cosmological constant $\Lambda$, the observable entropy of our universe will saturate at
\be
  S_{\rm dS} =  \pi m_p^2 H^{-2} = \frac{3\pi m_p^2}{\Lambda} \sim 10^{123},
\ee
where $m_p=1/\sqrt{G}$ is the Planck mass and $H$ is the Hubble parameter. Applying a similar formula during an inflationary phase, (a quasi-de Sitter space of much higher energy density),  $S_{\rm inf} \sim 10^{12}$, there has evidently been an enormous increase in entropy.  

In some cosmological models, however, equilibrium states that hold for arbitrarily long timescales can exist, and in this context significant \emph{downward} entropy excursions would, inevitably, occur. In fact, in models of eternal inflation (see, {\it e.g.},~\cite{Aguirre:2007p358,Guth:2007ng} for a review), these excursions may be crucial.  They are involved, for example, as intermediate states in the Coleman-de~Luccia or Hawking-Moss mechanisms by which transitions occur between metastable inflationary vacua, and are central to process in which the vacuum energy increases, as in stochastic eternal inflation or via the Lee-Weinberg mechanism.  They also describe, in eternal thermal spaces, the spontaneous emergence of structures such as black holes or Boltzmann Brains from empty space. (All of these processes are discussed in detail below.) 

It would therefore be useful to better understand the precise nature of these excursions, which are acknowledged but vaguely described in the cosmological literature. It is often imagined that these fluctuations are sudden and dramatic transitions, or that they do not admit any classical description. We will argue that cosmological fluctuations to lower-entropy states should be thought of as the {\em time reverse} of a -- generally smooth, or at least gradual -- natural evolution from a low-entropy state into equilibrium. Through this argument, we hope to provide physical insight, and spacetime diagrams, for a number of such processes.

We begin in Sec.~\ref{sec-piano} with a toy model (a piano in a box) that illustrates our main point,
that the time development of a downward fluctuation in entropy resembles the time-reversal of ordinary
evolution toward higher entropy.
In Sec.~\ref{sec-nongrav} we summarize our argument for the non-gravitational physics of a system that is either purely isolated or connected to a fixed-temperature heat bath.  In Sec.~\ref{sec-grav} we then describe our central assumption, which is that the statistical mechanics of certain eternal gravitating (and cosmological) systems can be treated similarly, to arrive at a prescription for understanding exponentially unlikely events representing a downward excursion in total (gravitational plus non-gravitational) entropy.  In Sec.~\ref{sec-applications} we apply this idea to eight 
 processes.  In Sec.~\ref{sec-discussion} we summarize, and comment on the difficult questions these ``miraculous" processes pose for any model in which they are important.

\section{The piano in a box: a parable}
\label{sec-piano}

The key argument we wish to explore in this paper can be illustrated by a simple example.
Consider an ice cube in a glass of water.  For thought-experiment purposes, imagine that
the glass of water is absolutely isolated from the rest of the universe, lasts for an infinitely long time, and
we ignore gravity.  Conventional thermodynamics predicts that the ice cube will melt, and in a
matter of several minutes we will have a somewhat colder glass of water.  But if we wait long enough,
of order a recurrence time $e^S$ where $S$ is the equilibrium entropy, statistical mechanics 
predicts that the ice cube will eventually re-form. If we were to see such a miraculous occurrence, the central claim of this paper is that  
the time evolution of the process of re-formation of the ice cube will, with high probability, be roughly
equivalent to the time-reversal of the process by which it originally melted.  
(For a related popular-level discussion see \cite{feth}, ch.~10.) The ice
cube will not suddenly reappear, but will gradually emerge over a matter of minutes via unmelting.
We would observe, therefore, a series of consecutive statistically unlikely events, rather than one
instantaneous very unlikely event.  The argument for this conclusion is based on conventional
statistical mechanics, with the novel ingredient that we impose a \emph{future} boundary condition --
an unmelted ice cube -- instead of a more conventional past boundary condition.

While this claim might seem reasonable in the context of a simple system like an ice cube in
water, it becomes more startling when we apply exactly the same logic to more complex systems.
We will ultimately want to apply it to the entire universe, so let's warm up by considering an
intermediate scale:  a grand piano.

Consider a  piano of mass $M_p = 5\times 10^5{\rm \,g}$ 
in a perfectly, completely and impenetrably sealed box of length $L=500{\rm \,cm} \sim 2.5\times 10^7 {\rm \,eV}^{-1}$
on each side, creating
a truly isolated system.\footnote{To avoid any other subtleties, we could replace the box walls by topological identifications of Minkowski space, to create a toroidal space.}
We also imagine that gravity is non-existent, so the piano is weightless, and, {\it e.g.}, black holes cannot form.
Finally, we posit that there are suppressed
operators that violate both baryon and lepton number, so that there are no conserved charges
other than energy-momentum inside the box. 

Let us first estimate the entropy of the piano's initial state.  (Order-of-magnitude precision will suffice
for this discussion.) It is made primarily of wood, which in turn
consists mostly of cellulose: long macromolecules made of strings of C$_6$H$_{10}$O$_5$,
with a typical molecular weight $\sim 10^5$.  The piano therefore 
has $\sim 3\times 10^{24}$ molecules, or $\sim 5$ moles.  For a very rough entropy estimate, the standard molar entropy $S_m^0$ of organic compounds 
in the solid state at room temperature 298~K is $\sim 10^2 {\rm\, J/(K\,mol)}$.  
This quantity goes up with molecular complexity, and cellulose is a very
large molecule, so we will estimate $S_m^0\sim 10^3 {\rm\, J/(K\,mol)} \sim 10^{26}  {\rm \, mol}^{-1}$.
(We set $k_B=1$ in the last expression, and will do so henceforth.)
The entropy of the piano at room temperature is therefore
\be
  S_P \sim 5 \times 10^{26}.
\ee

Consider now the actual evolution of the piano starting from its initial state and evolving for an indefinite time. The piano is metastable, and not in its highest entropy state; over time, through fluctuations 
it will evolve into a series of higher-entropy configurations.  First, thermal fluctuations will cause
individual molecules to break free of the larger structure, essentially sublimating into the
box.  The strings will likely break early, snapping periodically with a loud sound that punctuates eons of silence. 
Further decay releases chemical potential energy, slowly heating the box and in turn speeding the piano's decomposition.
 Eventually, the piano's structure will become so compromised that pieces will begin to break off and float away.
Over longer timescales, chemical and eventually nuclear reactions will alter the composition of the
original piano, first into a state of chemical equilibrium and ultimately into nuclear statistical
equilibrium. The box at that point will contain a gas of various ionized nuclei, electrons, photons, and neutrinos.  
If we imagine that there are GUT-scale interactions that violate baryon and lepton number, then on roughly the timescale of proton decay, $ \sim 10^{35}\,{\rm years}$,  the atoms will dissolve into photons and neutrinos, finally achieving a
high-entropy state.  

This gas of thermal particles filling the box represents the highest-entropy state of the system.
The total energy is $E_p = M_pc^2 \sim 3\times 10^{38}{\rm \, eV}$, while the energy density in a
relativistic gas is
\be
 \rho = \frac{\pi^2}{30}g_*T^4,
\ee 
where $g_* = g_{\rm bose} + (7/8)g_{\rm fermi}$ is the number of relativistic degrees of freedom at that temperature.  
For the temperatures relevant here, we assume that the relativistic degrees of freedom consist of one
photon and three species of massive neutrinos, for $g=7.25$.  Setting $E_p = \rho L^3$ gives
\be
 T \sim 10^4{\rm \, eV} \sim 10^8 {\rm \, K}.
\ee
The entropy density in a relativistic gas is 
\be
s= \frac{4\pi^2}{45}g_*T^3.
\ee
The total entropy of our piano-turned-gas is therefore
\be
S  \sim 5\times 10^{39}.
\label{pianogasentropy}
\ee

While the details of the evolution to higher entropy are interesting, in this paper we are concerned
with what happens afterward -- the rare fluctuations into lower-entropy states.  Since the system
has a finite entropy in thermal equilibrium and therefore a finite number of accessible states, the Poincar\'e recurrence
theorem guarantees that it will eventually cycle through all allowed configurations over a 
recurrence timescale of approximately
\be
  t_{\rm recur} \sim e^{S_{\rm max}} \sim e^{10^{39}}.
\ee
Given the magnitude of this number, the units are essentially irrelevant. Among the allowed states
is of course the original piano, so if we wait long enough the piano will reassemble
itself through purely random fluctuations.

What does the process of reassembly look like?  Since the initial configuration of the piano had
an entropy of order $10^{27}$, there are a huge number of microstates corresponding to that type
of configuration, and a correspondingly large number of evolutions that will lead to it.  However, we
can use statistical mechanics to estimate the most probable route the 
system takes. The answer is that the set of likely
routes to reassembly looks just like the time-reverse of the likely trajectories by which the piano originally relaxes to equilibrium.
The argument is fairly straightforward:  we use precisely the
same kind of reasoning by which we predict the evolution of the original piano toward equilibrium,
except that we now seek trajectories compatible with a \emph{future} boundary condition rather than
one in the past.  But since the dynamical equations themselves are essentially time-reversal invariant,
the sets of trajectories are isomorphic.

It's tempting to think of the process of piano re-assembly as a ``fluctuation'' that occurs
relatively quickly. But instead, the process is very gradual, 
consisting of a series of concatenated unlikely events rather than a single event. Over the course of
$\sim 10^{35}$\ years, baryon-number-violating process gradually create a net baryon and lepton 
number out of the gas of photons and neutrinos. These baryons assemble into nuclei, which 
gradually fluctuate into lighter organic elements. The resulting atoms assemble into the
form of cellulose and the other piano constituents, which in turn assemble into shards of wood and
ultimately into pieces of piano floating about. At some point the pieces painstakingly assemble into the form of a full piano, 
albeit one that looks somewhat worn, with broken strings. Given
more time, sudden events create sound waves in the room that focus onto a broken piano string, oscillating it until it whips into place and 
fuses the broken ends together. Finally,
 the wear on the piano gradually disappears, as remnant molecules join onto the piano to
complete the form it started with, complete with a final polish.

This story seem surprising not because the net result is unlikely, but because it consists of such a large
number of individually unlikely events.  Even if we grant that at some point a gas of photons and 
neutrinos could fluctuate a relatively large baryon number, it doesn't seem plausible that those baryons
would gradually convert into cellulose molecules, or that a collection of piano pieces would spontaneously
leap into the form of a full piano.

This clash with intuition comes about because we are unused to working with future boundary
conditions.  Given that a box of relativistic gas fluctuates into the form of broken piano pieces lying
on the floor of a box, it is extremely unlikely that they will go on to fluctuate into a piano.  
Of all the trajectories that start in equilibrium and end with shards of wood, only a tiny fraction
continue on to create the unbroken musical instrument.  But that is not the question we are asking; rather,
we are concerned with the form of typical trajectories that \emph{do} make it all the way to the piano.
Our claim is that most such trajectories consist of many fluctuations that are individually unlikely,
and given by the time-reversal of a low-entropy state evolving into a high-entropy one.

\section{Entropy fluctuations in non-gravitating systems}
\label{sec-nongrav}

In this section we provide a more detailed and careful justification for the claim that downward
fluctuations in entropy resemble the time-reversal of ordinary evolution toward increasing entropy, 
and extend it to more general circumstances. 
The reader who is already convinced of this can safely skip to Sec.~\ref{sec-grav},
where we apply this reasoning to cosmology.
		
\subsection{Evolution, entropy, and its increase and decrease}

Consider an idealized fully isolated system with a space of microstates $\Gamma$.  This may be either a continuous set $\{a\}$ or a discrete set $\{a_i\}, \ i=1...N$.  The discrete case could include finite-state models or points in a discretized classical phase space; the continuous case would include classical phase space. We discuss quantum mechanical systems separately in Sec.~\ref{sec:quantum}. Assume also that these states evolve via a reversible (unitary) operator $U_{t-t'}$ that takes each such state at time $t'$ into a state $U_{t-t'}a_i$ at $t$, such that $U_{t-t'}a_i=U_{t-t''}U_{t''-t'}a_i.$\footnote{In the discrete case time must also be discrete, or else the evolution operator written in terms of $\theta$-functions etc., so that it acts discretely.}

Now let us coarse-grain these microstates into macrostates $A_\alpha$, $\alpha=1...M$. Two microstates are said to be in the same macrostate if they are macroscopically indistinguishable, {\it i.e.} if they have identical values of certain specified macroscopic observables (up to some specified tolerance).  Classical states may be partitioned by simply assigning each state a single macrostate.  

The space of states comes equipped with a measure, which allows us to define the Boltzmann entropy of a macrostate. In the discrete case, the measure is provided by counting, and the volume $\Omega_\alpha$ is the number of elements of ${A}_\alpha$.  In classical mechanics on a $2n$-dimensional phase space we have the Liouville measure $\omega^n$, where $\omega = \sum_k dp_k\wedge dq^k$.  The corresponding volume of a macrostate $A_\alpha$ is $\Omega_\alpha = \int_{A_\alpha}\omega^n$. In either case the Boltzmann entropy is given by
\be
  S^B_\alpha = k_B \log \Omega_\alpha.
\ee
Note that the Boltzmann entropy is \emph{not} a measure of our knowledge of the system; it is a reflection of the coarse-graining.  Even if we know the microstate precisely, the Boltzmann entropy is objectively defined (and non-zero) once our coarse-graining is fixed.  In the (likely) event that a non-equilibrium microstate evolves into a macrostate with larger volume, $S^B$ goes up; in the (unlikely) event that it evolves into a macrostate with smaller volume, it goes down.
  
Now let us imagine we have some probability distribution for the microstates at each time (stemming either from ignorance or because we are describing an ensemble or collection of such systems).  For a finite set of states $a_i$, we can assign probabilities $p_i$, $\sum_i p_i=1$; for states continuously labeled by $x$ (as in a classical phase space), let $\rho(x)$ be the probability density with $\int dx \rho(x)=1$.  We may then also define a statistical entropy,\footnote{This is also the Gibbs entropy, if the normalization is chosen to agree with thermodynamic entropy. Note also that in the continuous case the entropy must be taken as a continuum limit of a discretized case, and will be defined only to an an infinite additive constant that is independent of the probability distribution.}  
\be
  S^S = -\sum_i p_i \log p_i.
\ee
In the continuous case this becomes $S^S = -\int dx \rho(x)\log\rho(x)$.

\begin{figure*}[htb]
\includegraphics[width=15 cm]{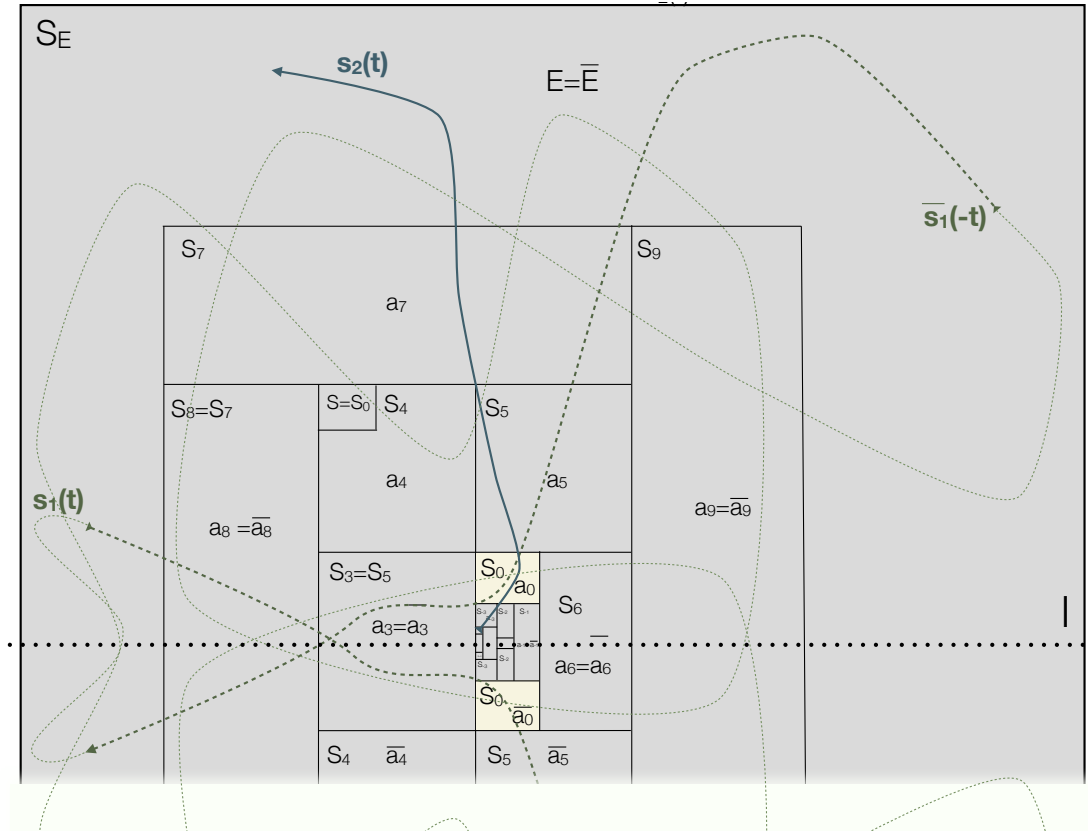}
\caption{
A schematic diagram of a general space of states $\Gamma$.  The space is partitioned into macrostates $A_\alpha, \alpha=0...9$, with Boltzmann entropies $S_\alpha$, as well as an equilibrium macrostate $E$ with entropy $S_E$.  (Relative phase space volume are vastly diminished relative to any realistic situation). Microstate trajectories $a_1(t) \equiv U_{t-t_0}a_1$ and $a_2(t)$ are shown, where both `start' in $A_0$ at $t=t_0$ ($a_1, a_2 \in A_0$) and evolve according to the unitary operator $U$.  Trajectory 2 represents a system `prepared' in a very low-entropy macrostate that evolves into $A_0$ and thereafter into macrostates of ever-increasing entropy; trajectory 1 wanders by chance into $A_0$ from equilibrium, tracing out a path that is symmetric in entropy, but asymmetric in the macrostates traversed. The system features an involution containing time-reversal, which maps $a_i(t) \rightarrow \bar a_i(-t)$ with $\bar{\bar{a}}_i(-t)= a_i(t)$; the action of this map on the states is depicted by a reflection across the `I' axis.  Macrostates $\bar A_\alpha$ are defined by the action of time-reversal on their component microstates: $\bar a_i \in \bar A_\alpha \iff  a_i \in A_\alpha$.  The micro-trajectories have probabilities $p_i(t)$ determined by $p_i(t_0)$, and these yield also probabilities $P_\alpha(t)$ for the macro-trajectories.  The prescription for obtaining $P_\alpha(t)$ for states entering $A_0$ is to consider micro-trajectories leaving $\bar A_0$, because there is exactly one of these for each micro-trajectory entering $A_0$.
  \label{fig-phasespace}
}
\end{figure*}

The individual probabilities $p_i$ can evolve in time if we allow for a re-labeling of microstates. The probability of a state $i$ at $t$ is just given by the probability at time $t_0$ of the state $j(i)$ from which state $i$ evolved: $p_i(t) = p_{j(i)}(t_0)$, where $a_i = U_{t-t_0}a_{j(i)}$. The original partitioning and assignment of probabilities at $t_0$, together with the unitary evolution of microstates, is sufficient to determine the evolution of the micro-probabilities $p_i(t)$. Since the microstates evolve unitarily, the evolution of the micro-probabilities will be time-reversible. See Fig.~\ref{fig-phasespace} for a schematic diagram depicting ideas of this section.

In discretized or continuous classical Hamiltonian systems $S^S$ will be constant in time. In the discrete case, this is because the set of assignments of $p_i$'s to states is simply re-arranged under time evolution. In the continuous case, it is a consequence of Liouville's equation $d\rho/dt=0$, where $\rho$ is the phase space density.

The evolution of the microstates (and their attendant probabilities), along with coarse-grained macrostates $A_\alpha$, also defines a set of probabilities and entropies over the macrostates. The discrete case is simplest: we assign probabilities $P_\alpha(t) = \sum_{i\in I_\alpha} p_i(t)$ to our macrostates, as well as corresponding conditional probabilities $p_{i|\alpha}$ defined by $p_i = p_{i|\alpha}P_\alpha$.  
(It may seem that this equation needs a sum over $\alpha$, but only the macrostate to which $a_i$ belongs
actually contributes.)  

Imagine that our system is known at time $t_0$ to be in some far-from-equilibrium macrostate $A_{\alpha_0}\equiv A_0$, so that $P_\alpha(t_0) = \delta_{\alpha,\alpha_0}$. In the discrete classical case, this macrostate corresponds to microstates $a_i$, $i\in {A}_{\alpha_0}$, with probabilities $p_{i|\alpha_0}$, each of which will evolve in time. The evolution of $P_\alpha$ then follows from that of the microstates, with each evolved microstate $i$ contributing probability to the appropriate macrostate: 
$$
P_\alpha(t) = \sum_{i\in {A}_\alpha} p_{i}(t).
$$
The character of this evolution of $P_\alpha$ depends, in general, both upon the evolution of the microstates and on the choice of macrostates.  Our argument does not really depend on how this evolution proceeds in detail, but it is interesting to discuss the relation between this evolution and that of (various types of) entropy.

Because we can associate a statistical entropy with any probability distribution, in addition to the ``Boltzmann" entropy $S^B_\alpha$ of a macrostate and the ``Gibbs" entropy $S^S = -\sum_i p_i\log p_i$, we can also define
$S_\alpha \equiv -\sum_i p_{i|\alpha}\log p_{i|\alpha}$ (the statistical entropy of a given macrostate) and also $S^A \equiv  -\sum_\alpha P_\alpha \log P_\alpha$ (a coarse-grained statistical entropy).  These four entropies are related: if we assume that all microstates are equally probable when a macrostate is specified, then $S_\alpha = S^B_\alpha$ up to a normalization constant; it is also straightforward to derive that
\begin{equation}
S^S = S^A + \sum_\alpha P_\alpha S_\alpha.
\label{eq-srelations}
\end{equation}

What happens to these quantities as our initial state $A_0$ evolves?  One basic result follows if we assume that the partitioning into macrostates is decoupled from the details of the microstate probabilities $p_i$ and the evolution $U$.  
In this case, trajectories leaving one macrostate $A_{\alpha_0}$ for another will tend to enter neighboring macrostates with large numbers of component microstates, i.e. large Boltzmann entropy. This is, of course, a qualitative statement of the Second Law.  
The other entropies area a bit more subtle because they can change discontinuously depending on our knowledge of the system.  To the system originally in macrostate $A_0$ we would assign $S^A=0$ and $S^S=\log\Omega_0$ (perfect macro-knowledge, and perfect micro-ignorance).  As the system evolves unitarily, $S^A$ generally increases as the probability spreads to multiple macrostates, while $S^S$ is fixed as discussed above. Thus the increase in $S^A$ comes at the expense of the second term in Eq.~\ref{eq-srelations}, and because the $S_\alpha$'s are non-negative, under such evolution it can never exceed $\log\Omega_0$. Now imagine at some time observing the system and determining it to be in some higher-entropy macrostate $A_1$ (either dispelling our ignorance or choosing a member out of some ensemble.)  $S^A$ again drops to zero, $P_\alpha=\delta_{\alpha1}$, and $S^S = \log\Omega_1=S^B_1$.  The entropy has increased as per the second law just as before, and quite possibly by a large amount due to our `throwing out' the micro-information we had from our knowledge that the system had evolved from $A_0$.  Note, however, that if we were to continually measure the system macroscopically, we would observe a steady entropy increase.  Entropy {\em decrease} is rather similar. $S_S$ starts high, and by unitary evolution never changes.  However, at any given time a measurement of the system might reveal it to be in a macrostate with small $S_\alpha$, so we say that entropy has ``suddenly" decreased, although the physical evolution has been smooth, and constant measurement of the system would have revealed a smoothly decreasing entropy. The only real difference is that this entropy change should be exponentially rare, rather than generic.

Note that the difficulty in understanding the thermodynamic arrow of time is not in explaining why entropy increases in one `forward' time direction from $t_0$, but rather in explaining why it could {\em decrease} in the other `backward' time direction.  The assumption that entropy was, in fact, lower in the `past' is often termed the `past hypothesis' (see~\cite{albert}), and indicates that systems in the real world are {\em not}, in fact, equally likely to be in any microstate compatible with their observed macrostate, but rather are in one of a tiny subset of microstates that would evolve toward lower entropy in the decreasing time direction.  We shall assume, however, that this tiny subset is nonetheless an accurate sample of all of the states in $A_0$, so that given a macrostate at time $t_0$, its evolution toward greater times does not depend on whether it had lower entropy to the past.  This assumption -- that a ``prepared" system will evolve toward equilibrium in the same manner as a fluctuation -- is known as ``Onsager's regression hypothesis"~\cite{PhysRev.38.2265} and is an assumption in the fluctuation-dissipation theorem. There, it is used to explain prepared system in terms of fluctuations.  Here, we will use what we think we know about the evolution of prepared systems to make inferences about the evolution of fluctuations.

We may now ask the central question of this paper: if a system evolves from equilibrium to a state of low entropy, what is the (most probable) sequence of macrostates by which it does so?

\subsection{A non-gravitational system of fixed energy}

As a first, relatively well-defined, case, consider an isolated non-gravitational system of fixed energy $E_0$ that starts in a non-equilibrium macrostate $A_0\equiv A_{\alpha_0}$, then evolves to equilibrium.  
Statistically, equilibrium is described by the microcanonical ensemble, in which we assume that all microstates of the given energy $E_0$ (to within a tolerance $\delta E$) are equally probable. Thus $P_\alpha \propto \Omega_\alpha \delta(E-E_0)$. 
However, under our assumptions,\footnote{Or even more general ones: the Poincar\'e recurrence theorem just requires a one-to-one measure-preserving evolution map from a state space of finite measure onto itself, that maps a subset of finite measure onto another subset of finite measure.  It thus applies to both classical dynamics and standard quantum mechanics; see, {\it e.g.}~\cite{Schulman:1997p3370} for a simple and general proof, and~\cite{Dyson:2002pf} for a summary of an alternative proof in the quantum case.}
the Poincar\'e recurrence theorem guarantees that our system returns arbitrarily close to any given microstate, and thus necessarily eventually returns to macrostate $A_0$. What macrostates preceded this state?  

Consider the microstates $a_i$ making up $A_0$.  Under unitary evolution each of these traces out a trajectory $a_i(t) = U_{t-t_0}a_i$ through state-space. We now assume that the physics of the system in question is ``time reversible," meaning that for each solution to the equations of motion, there is a time-reversed solution that is also a solution. (In quantum field theory one needs CPT.) More precisely, there is an involution $T$ that maps the set of possible trajectories $a_i(t)$ into another set, $T:{a_i}(t) \rightarrow \bar{a}_{i}(-t)$, such that $T^2 = {\bf 1}$.  In classical $N$-particle mechanics of the phase space $\{\vec x_i,\vec p_i\}, i=1...N$, $T$ maps $\vec x_i(t) \rightarrow \vec x_i(-t)$ and $\vec p_i(-t) \rightarrow -\vec p_i(-t)$.
Finally, consider the set of states $\bar a_i(t_0), i\in {A}_{\alpha_0}$, constituting the ``time reverse" of the microstates making up $A_0$.  We shall assume that these correspond to some macrostate $\bar A_0$, i.e. that time-reversal maps macrostates to macrostates. If $\bar A_0 = A_0$, we shall call this macrostate a ``bounce state".  

With all this defined, we can now ask what macrostates preceded $A_0$ if it is a fluctuation from equilibrium at time $t_0$, or more precisely, what is the time-dependent macrostate probability distribution $P_\alpha(t)$ {\em given} that $P_\alpha(t_0) =  \delta_{\alpha,\alpha_0}$?  The key assumption we will use is that at $t_0$, all microstates $i\in {A}_{\alpha_0}$ are equally probable.\footnote{Again, this would {\em not} be true for an instance of $A_0$ that we ``prepare," or somehow know evolved from a lower entropy macrostate.}

 Consider the set of trajectories $a_i(t)$, where $a_i(t_0) \in A_0$.  Then the set $\left\{\bar a_i(t_0)\right\}$ is just the set of microstates of $\bar A_0$.  If we accord equal probabilities to each of these, we know the time-dependent probabilities $P_\alpha(t)$ describing the macroscopic trajectories away from this state (in increasing $t$), using just the same previous reasoning as to how $A_0$ evolves.  The $P_\alpha$'s then just describe the ``standard" evolution from the state $\bar A_0$ to equilibrium.  Now consider the time-reverse of this evolution (by time-reversing each macrostate). This gives exactly the same probabilities (since they are just proportional to the fraction of microstates in each macrostate), now governing the macro-trajectories that lead {\em to} $A_0$, and such that each microstate in $A_0$ is equally probable. It is thus exactly the set of probabilities we sought over the macro-trajectories leading up to our fluctuation out of equilibrium.
  
In summary: for a unitary system with a state space of finite measure and time-symmetric (meaning a symmetry including time-reversal) dynamics over that state space, fluctuations from equilibrium to any given low-entropy macrostate $A_{\alpha_0}$ will eventually occur. The probability distribution of macroscopic histories leading {\em from} equilibrium {\em to} $A_{\alpha_0}$ is the time-reverse of the PD leading {\em to} equilibrium {\em  from} the time-reversed macrostate $\bar A_{\alpha_0}$.

Applying this to, say, a piano moving to the right, the time-reversed macrostate would be a piano (technically, made of antimatter) moving to the left.  This piano would gradually decay over many aeons as described qualitatively in Sec.~\ref{sec-piano}.  There would be many possibilities for how this might happen in more detail (which leg detaches first, etc.), including some incredibly rare cases ({\it e.g.}, in which the piano spontaneously combusts).  Time-reversing this set of macroscopic histories yields the probability distribution for histories by which the (right-moving) piano would fluctuate from equilibrium.  A tiny fraction might contain sudden downward jumps in entropy (an `uncombusting piano'), but the vast majority would simply appear to be the detailed time-reverse of standard evolution toward maximal entropy.

\subsection{Thermal systems in equilibrium with an environment}

Let us now consider a finite thermal system that is coupled with an environment, such that the system+environment (S+E) is isolated, and described by the microcanonical ensemble of fixed energy $E$.  Our system is then described by the canonical ensemble (if it can exchange energy) or grand canonical ensemble (if it can also exchange particles), with a Boltzmann probability distribution $p_i$ (over a new state space $a_i$ including all energies) that results from the different ways in which the system and environment can share the fixed energy $E$.  

In terms of our question, there is actually very little change, given an appropriate coarse-graining of the microstates of the S+E.  Define ${\cal A}_\alpha$ to be macrostates of S+E such a microstate of S+E with index $i$ is in ${\cal A}_\alpha$ if and only if when we restrict to the system, the system is in macrostate $A_\alpha$.  That is, we use the coarse-graining on the system to induce one on S+E, which ignores all details about S+E that are not relevant to determining which macrostate the system is in.  Having this partition, and waiting for the system to attain macrostate $A_0$, we then know that S+E is in macrostate ${\cal A}_0$.  We can then -- in precisely the manner of the previous section -- time-reverse ${\cal A}_0$ into $\bar {\cal A}_0$, calculate the probability distribution over macrostates evolving away from $\bar {\cal A}_0$, to equilibrium, then time-reverse these macrostates to find those leading from equilibrium up to ${\cal A}_0$, and accord the same probability distribution to those.

These probabilities will imply that a system measured with some energy $E_0$ may at a later time be measured with an energy $E_1\neq E_0$.  But recapitulating the discussion of entropy earlier, this does not mean that anything sudden has necessarily occurred.  We can imagine measuring the macroscopic properties of the system continuously, so that it traces out some well-defined macro-evolution with fluctuating energy due to an external perturbation.  Note also that because fluctuations in the entropy of S+E are exponentially suppressed, the fluctuations in which the system's entropy decreases by a given amount are exponentially dominated by histories of S+E in which the environment's entropy is decreased as little as possible.

We can expect the behavior of a general thermal system to be similar, independent of the details of the bath to which it is coupled. 
 Indeed, this behavior seems quite general for systems with time-reversible dynamics, including systems with random perturbations in their evolution, as long as the statistical behavior of the random perturbations do not distinguish a time-direction.\footnote{For example, this is true in Langevin systems with Gaussain noise; see Sec.~\ref{sec:stochastic}.}
 
\subsection{Quantum mechanics}\label{sec:quantum}
	
Now let's consider the quantum case. The first step is to choose an appropriate coarse-graining
on the space of states, which is a more subtle task than it is in classical mechanics.  Following
von~Neumann \cite{springerlink:10.1140/epjh/e2010-00008-5,springerlink:10.1140/epjh/e2010-00007-7}, 
we decompose the entire Hilbert space $\calh$ into a set of mutually orthogonal subspaces representing
macrostates:
\be
  \calh = \bigoplus_A \calh_A.
  \label{quantummacro}
\ee
We denote the dimensionality of each subspace by $d_A = {\rm dim}(\calh_A)$.
von~Neumann argued that this decomposition could be sensibly defined by ``rounding'' a set of 
macroscopic quantum-mechanical observables to obtain a family of commuting operators 
that approximate classical observables.    The macrostate subspaces $\calh_A$ are then spanned by the
simultaneous eigenvectors of these observables within a prescribed range of eigenvalues.  The details
of this procedure are not germane to this paper; we will only rely on the existence of the 
decomposition (\ref{quantummacro}).

We can describe a quantum system in the macrostate $\calh_A$ specified by certain macroscopic observables 
using a density matrix given by an equal weight over an orthonormal basis of the subspace:
\be
  \rho_A = \frac{1}{d_A}\sum_{\alpha} |\phi_\alpha\rangle\langle \phi_\alpha|.
\ee
The precise form of this density operator will not be crucial in what follows; all that matters is that we
can represent our macroscopic states by specified mixed states, and that macrostates go to macrostates
under time reversal. The corresponding von~Neumann entropy is
\be
  S_A = \tr(\rho_A \log \rho_A) = d_A.
\ee
We can then phrase our prescription in the following way:  given that the system starts in a 
high-entropy mixed state $\rho_b$, its evolution to a lower-entropy mixed state $\rho_a$ is given
by the time reverse of the conventional evolution from $\rho_a$ to $\rho_b$.  We clarify the precise
operational meaning of this statement below.

Our problem is one of a future boundary condition, specified by a certain density operator.
The problem of future boundary conditions in quantum mechanics was studied by Aharonov, Bergmann, and
Lebowitz (ABL) \cite{PhysRev.134.B1410,aharonovrohrlich}.  In the Schr\"odinger picture, 
imagine that we start with a state $|a\rangle$ at time $t-\Delta t$, evolve it for a time $\Delta t$, and then
perform an observation.  The probability that the system is observed to be in a state $|c_n\rangle$
at time $t$ is given by
\be
  P(c_n, t|a, t-\Delta t) = |\langle c_n|U_{\Delta t}|a\rangle|^2,
  \label{born}
\ee
where $U_{\Delta t}$ is the unitary operator that evolves the state for a time $\Delta t$.
(We assume that the Hamiltonian is time-independent, so that the actual value of $t$ is irrelevant; this
assumption may be relaxed at the cost of more cumbersome notation.)
But now imagine that we post-select as well as pre-select: we demand that the system is in a state $|a\rangle$
at time $t-\Delta t_1$, but also that it is in a state $|b\rangle$ at a time $t+\Delta t_2$ that is \emph{after}
the observation is made.  ABL showed that the probability that the system is in a state $|c_n\rangle$
at an intermediate time $t$ is given by
\be
  P(c_n, t|a, t-\Delta t_1; b, t+\Delta t_2) = \frac{|\langle b|U_{\Delta t_2}|c_n\rangle\langle c_n|U_{\Delta t_1}|a\rangle|^2}
  {\sum_m |\langle b|U_{\Delta t_2}|c_m\rangle\langle c_m|U_{\Delta t_1}|a\rangle|^2}.
  \label{abl}
\ee
Here the $|c_m\rangle$ are a complete set of intermediate states.  The time-symmetry of the ABL formula
demonstrates that the apparent irreversibility in the ``collapse of the wave function'' is not an inherent
feature of quantum dynamics, but arises from the fact that we typically impose past boundary conditions
but not future ones (which arises in turn because of the thermodynamic arrow of time).  Note that
the operator $U_{\Delta t_2}$ can equally well be thought of as evolving $|c_n\rangle$ forward in time
by $\Delta t_2$, or as evolving $\langle b|$ backward in time by an equal amount.

Now consider the case where our boundary conditions are given by density matrices rather than
pure states. A density matrix evolves according to
\be
  \rho(t + \Delta t) = U_{\Delta t} \rho(t) U_{\Delta t}^\dagger.
\ee
Define the projection operator associated with an observation as
\be
  \Pi_n = |c_n\rangle\langle c_n|.
\ee
(This formula assumes nondegenerate eigenvalues, another restriction that is easily loosened.)
If we start in mixed state $\rho_a$ and evolve for time $\Delta t$,
the probability of then observing the state $|c_n\rangle$ is given by
\be
  P(c_n, t|\rho, t-\Delta t) = \tr[\Pi_nU_{\Delta t} \rho U_{\Delta t}^\dagger].
\ee
To implement a future boundary condition,
we would like to generalize the ABL formula by asking what the probability of measuring $|c_n\rangle$ is,
given that our starting state a time $\Delta t_1$ earlier is $\rho_a$ and our final state  a time
$\Delta t_2$ later is $\rho_b$.  The answer can be written
\be
    P(c_n, t|\rho_a, t-\Delta t_1; \rho_b, t+\Delta t_2) 
    = \frac{\tr\left[ \Pi_{n}U_{\Delta t_1}^\dagger \rho_b U_{\Delta t_2}\Pi_{n} 
      U_{\Delta t_1}\rho_a U_{\Delta t_1}^\dagger\right]}
  {\sum_m \tr\left[  \Pi_{m}U_{\Delta t_1}^\dagger \rho_b U_{\Delta t_2}\Pi_{m} 
      U_{\Delta t_1}\rho_a U_{\Delta t_1}^\dagger\right]}.
    \label{rho-abl}
\ee
This equation can be straightforwardly derived by following the logic presented in \cite{0305-4470-24-10-018} for the pure-state case (see also~ \cite{GellMann:1991ck}.)\footnote{Ref. \cite{PhysRevA.57.2251} generalizes the ABL formula to the case of multiple final states.}
It is also easy to allow for multiple measurements at different times by inserting the 
appropriate projection and evolution operators.

Now we can ask about the time-reversed situation.  We introduce a time-reversal operator $T$ that
is an involution on Hilbert space, $T = T^{-1}$. In quantum field theory, we would need CPT; the details 
of the argument below are unchanged in this case. Time reversal acts on operators via conjugation; we denote
the time-reversed projection and density operators by
\be
  \Pi_n^T = T\Pi_nT^{-1}\,, \quad \rho^T = T\rho T^{-1} .
\ee
Time reversal is an antiunitary operator, $Ti = -iT$, and we assume that it commutes with the Hamiltonian,
$HT = TH$. On the unitary time-evolution operator $U_{\Delta t}=\exp(-iH\Delta t)$ we therefore have
\be
  TU_{\Delta t}T^{-1} = U_{-\Delta t}TT^{-1} = U_{-\Delta t} = U_{\Delta t}^\dagger.
\ee
We also assume as before that the time-reversal of two quantum states in the same macrostate
are in the same time-reversed macrostate; {\it i.e.} that time reversal commutes with coarse-graining.

Now take the numerator of (\ref{rho-abl}) and insert $T^{-1}T={\bf 1}$ between each operator.  Using the
above relations and the cyclic property of the trace, we get
\be
  \tr\left[ \Pi_{n}U_{\Delta t_1}^\dagger \rho_b U_{\Delta t_2}\Pi_{n} 
      U_{\Delta t_1}\rho_a U_{\Delta t_1}^\dagger\right]
  = \tr\left[\Pi_{n}^T U_{\Delta t_1}^\dagger\rho_a^T U_{\Delta t_1}
    \Pi_{n}^TU_{\Delta t_2} \rho_b^T U_{\Delta t_1}^\dagger \right].
\ee
An analogous expression holds for the denominator of (\ref{rho-abl}).  The interpretation of the
right-hand side of this equation is that we start in a mixed state $\rho_b^T$, evolve forward in time
by $\Delta t_2$, make an observation using $\Pi_n^T$, then evolve forward in time by $\Delta t_1$
and post-select on the mixed state $\rho_a^T$.  In other words, precisely the time-reverse of our
original situation.  We therefore have
\be
  P(c_n^T, t|\rho_b^T, t-\Delta t_2; \rho_a^T, t+\Delta t_1) 
  =P(c_n, t|\rho_a, t-\Delta t_1; \rho_b, t+\Delta t_2) .
\ee

This is the main result of this subsection.  The time-development of a quantum system fluctuating
from a high-entropy state $\rho_b^T$ to a low-entropy state $\rho_a^T$, as observed by a sequence
of measurements at intermediate times, is statistically equal to the time-reverse of the development
of a system from a low-entropy state $\rho_a$ to a high-entropy state $\rho_b$.  We will not use
the explicit formulas of this section in the rest of this paper, but they provide a justification for our
reasoning.

\section{Thermal systems, gravity, and cosmology}
\label{sec-grav}

Significant downward entropy excursions in macroscopic systems are so rare that they are non-ignorable only in systems that can exist for vast times, requiring consideration of both cosmology and gravity.  While we are far from understanding the full quantum theory of gravity, we have surprisingly strong (theoretical) reason to attribute thermal properties to certain spacetimes, including black holes and de Sitter (dS) space.  Other spaces -- including Minkowski and Anti de Sitter (AdS) -- are not ``intrinsically" thermal, but can host thermal fields.  In each of these systems, it has been argued -- with different degrees of rigor and as summarized below -- that there is a well defined equilibrium state as well as fluctuations away from it.  Insofar as this is the case, and physics -- including gravity -- satisfies the assumptions of 1) time-reversal (CPT) invariance and 2) democracy of microstates corresponding to each macrostate, the statements we have made in previous sections about fluctuations away from equilibrium will hold. 

\subsection{Thermal Anti de Sitter space}

We start by describing what is perhaps the most rigorous example of a gravitational system that can attain thermodynamic equilibrium: thermal Anti-de Sitter (AdS) space. As was first shown by Hawking and Page~\cite{Hawking:1982dh}, black holes with horizon radius exceeding the AdS radius $\ell_{AdS} = 3 |\Lambda|^{-1/2}$ (where $\Lambda$ is a negative cosmological constant) can come into thermal equilibrium with their own Hawking radiation. This occurs because the timelike boundary of  AdS space acts like a box containing the black hole, and reflecting any radiation that it emits. In principle, through the AdS/CFT correspondence, there is an alternate description of the full gravitational and matter degrees of freedom for asymptotically AdS systems in terms of a non-gravitational conformal field theory (see {\it e.g.}~\cite{Aharony:1999ti} for a review). Famously, these large black holes have an alternate description as a finite temperature field theory. This directly reduces the gravitational case to the standard non-gravitational case, in which we would expect both time-reversal (CPT) invariance of the microphysical laws and (arguably) a democracy of microstates. The description of fluctuations away from this equilibrium state, on both sides of the duality, should therefore follow from our prescription. In the following sections, we will draw directly on this example. However, we also rely on it as a general proof-of-principle that an understanding of fluctuations in non-gravitational systems can directly inform us on the nature of fluctuations in gravitational systems.

\subsection{de Sitter space as a thermal system}

The most relevant equilibrium system for cosmology is de Sitter (dS) space. The presence of an event horizon imbues dS space with ``thermal" characteristics. The most concrete manifestation of this is the thermal spectrum of particles~\cite{Gibbons:1977mu} of temperature $T = 2 \pi H$ (where $H$ is the Hubble constant and $H^{-1}$ the radius of the cosmological horizon) that any comoving detector will observe. In analogy with the thermodynamics of black holes, this implies an entropy for a horizon volume of dS space
equal to $S = A  / 4 G_N = \pi m_p^2 H^{-2}$. Semi-classically it appears that dS has an equilibrium, and just as for any system at either fixed temperature or energy, there will be fluctuations. However, the fundamental degrees of freedom underlying dS space are unknown. In light of this, how might we understand the statistical mechanics of dS space?

There are a number of reasons why a theory of dS space has been elusive. Because of the existence of a cosmological event horizon, there is a distinction between the amount of information necessary to specify the experiences of any one observer, and the information necessary to specify processes happening globally in the whole spacetime. The cosmological horizon makes it impossible to define an S-Matrix~\cite{Bousso:2004tv}, and places a fundamental limit on the accuracy of any measurement~\cite{Banks:2002wr}. Therefore, the amount of information, and information processing power, available to any single observer is finite and fundamentally bounded. Globally, this is not so: patches of the universe go out of causal contact, and seemingly become independent. Therefore, it would seem to take an infinite amount of information to specify what is happening globally in dS space. In this global context, we might then think of the horizon entropy as a form of entanglement entropy coming from tracing out the unobservable degrees of freedom in other hubble patches. 

There are a number of ways to interpret what all this means for the fundamental theory of dS space. In analogy with black holes, it has been suggested that a form of complementarity~\cite{Dyson:2002nt} exists in dS space. This implies that any one observer can describe the entire ``system," which should be thought of as corresponding to the region of spacetime accessible to this observer. Complementarity plus the bound on the information accessible to any one observer also implies that dS can be described by a theory with a finite number of degrees of freedom, $\mathcal{N} \sim e^{S_{dS}} = e^{\pi m_p^2 H^2}$ (see {\it e.g.}~\cite{Banks:2000fe,Bousso:2000nf} and recent attempts to perform microstate counting in string theory constructions~\cite{Dong:2010pm}; for a counter example to this claim, see~\cite{Bousso:2002fi}). In contrast, a more ``global" picture of de Sitter can be found in various attempts to define a dS/CFT correspondence~\cite{Strominger:2001pn} (or some boundary measure on an eternally inflating universe~\cite{Garriga:2009hy,Sekino:2009kv}). It is not clear how to reconcile these interpretations, and we have little to add to the discussion surrounding these two points of view.

However, since we are mainly concerned with the { \em observable} behavior of fluctuations, the differences in these viewpoints may not be crucial. In order to apply the arguments given above for fluctuations in non-gravitational systems, we will assume that whatever the fundamental degrees of freedom underlying dS space are, they evolve with respect to a CPT invariant set of laws. If dS space has a finite number of degrees of freedom, then a Hamiltonian governing their evolution has been postulated by Banks~\cite{Banks:2005bm}. A more global picture may admit a description in terms of a dual field theory. In either case, it seems reasonable to assume that CPT-invariance holds. The main difference between these two viewpoints may then resemble the difference between the canonical ensemble (corresponding to a ``global" picture in which the observable portion is coupled to a ``heat bath" representing everything else) and microcanonical ensemble (corresponding to a ``local" picture in which the region accessible to a single observer is the whole system). With the additional assumption of democracy of accessible microstates, we may apply the arguments of the previous sections to fluctuations from equilibrium in dS space.

\subsection{The prescription}
\label{sec-prescription}

The ``prescription" to be employed, which combines the arguments of Sec.~\ref{sec-nongrav} with the assumptions about the physics of dS and AdS space from this section, will be as follows.  

First, consider the causal diamond (the intersection of the causal past and causal future) of a past- and future-inextendable timelike worldline (the ``observer"), remaining when possible at the center of spherical symmetry.  The ``system" will be taken to be fields (including the metric) on a progression of spacelike surfaces that are orthogonal to the observer's worldline and that together foliate the spacetime within the causal diamond. We will adopt the philosophy that events outside of the causal diamond can be ignored, or at least are irrelevant to experiments that any observer can in principle perform.

In pure AdS, the causal diamond of an observer can enclose the entire global spacetime. In pure dS, the causal diamond encompasses at most a finite four-volume of order $H^{-4}$. For metastable dS, the accessible four-volume is observer-dependent, going to infinity for observers entering a region with zero cosmological constant (these are the so-called ``census takers" \cite{Susskind:2007pv,Sekino:2009kv,Harlow:2010my}). In such cases, it is unclear how to define the ``system," (although see~\cite{Freivogel:2006xu,Sekino:2009kv}) but we will assume that for any observer never entering such a region, it makes sense to have a description in terms of a causal diamond of finite four-volume. 

We define some (macro)state $A_0$ of this system, the natural evolution of which will be towards equilibrium dS or thermal AdS.  For simplicity, we will in many cases consider states that are spherically symmetric and momentarily static, so that they are ``bounce" states in the terminology of Sec.~\ref{sec-nongrav}.  

We then ask: if the system fluctuates from equilibrium to $A_0$, then what is the most likely evolution by which it does so?  The argument of Sec.~\ref{sec-nongrav} provides the prescription: we take the CPT-conjugate of $A_0 $ (which due to our ``bounce" assumption generally does nothing), then let this initial condition evolve to equilibrium to trace out a set of possible macroscopic histories of the system, and define their probability $P_\alpha(t)$. The probability for observing some history leading up to $A_0$ is then given by the CPT conjugate of the probability distribution over histories evolving away from $A_0$. We will take this prescription to apply equally well to metastable equilibria that can live long enough for the desired fluctuation to occur.

Note that our prescription singles out a time foliation (or group of foliations) in which the system is in equilibrium. This naturally follows from our restriction to physics inside of a causal diamond; singling out an observer introduces a preferred frame. We comment briefly on the implications of this below.

\section{Applications}
\label{sec-applications}

In this section we will apply our prescription to a set of processes that can accurately be considered as downward entropy fluctuations in a thermal system. To keep this paper of reasonable length and because many of these processes are well-described in the literature,  we summarize these processes fairly succinctly (with references to fuller accounts elsewhere), emphasizing what our prescription adds or clarifies.

Note that several of the processes we discuss, such as decay of the false vacuum via Coleman-de~Luccia
instantons, ultimately represent evolution toward higher-entropy states.  However, they do so by passing
through intermediate configurations of lower entropy.  Our analysis is therefore applicable to these situations,
and helps to illuminate the time development of such transitions.

\subsection{The ``thermalon"}

\begin{figure*}[htb]
\includegraphics[width=12 cm]{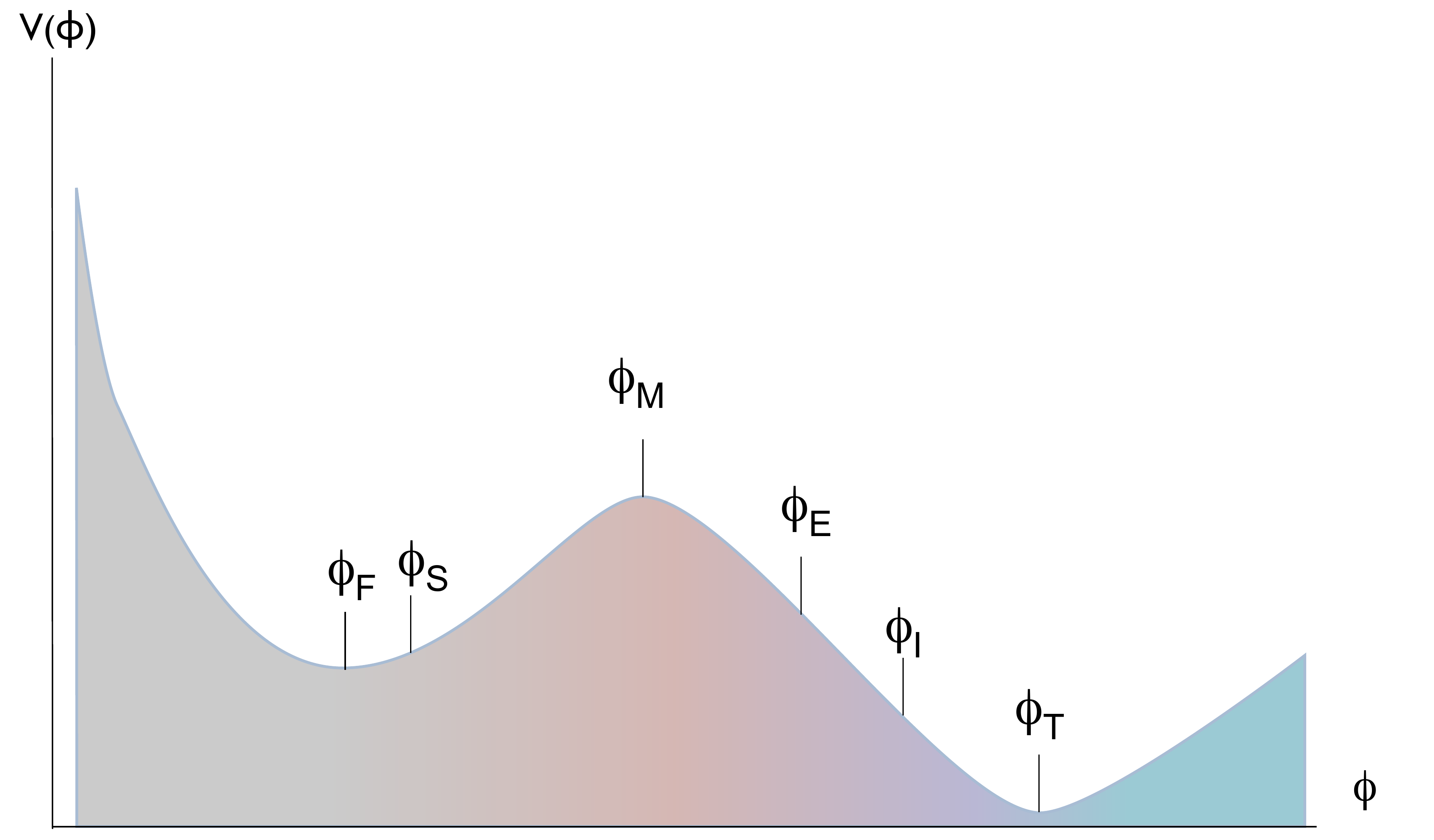}
\caption{
Schematic scalar potential for the thermalon, Coleman-de~Luccia, Hawking-Moss, and Lee-Weinberg processes (see text).  In reality not all processes can occur  in the same potential, but a single figure can represent the necessary qualitative features. See subsequent figures for the meanings of the labeled field values in those various contexts.
\label{fig-doublewell}
}
\end{figure*}

The first process we shall discuss is the creation of a local static scalar field configuration for a single field in the double-well potential $V(\phi)$ depicted in Fig.~\ref{fig-doublewell} (which will also serve for several further processes). For now, let us work in the absence of gravity at finite temperature. If the field is everywhere in the metastable false vacuum, a pocket of the true vacuum can develop via the formation of a {\em static} field configuration at the threshold between expansion and re-collapse~\cite{Linde:1980tt}. This corresponds to an O(3)-invariant instanton that, for consistency in terminology, we refer to as the ``thermalon", the name given to its gravitational cousin. (See~\cite{Garriga:2004p3391} and references therein.)

How would the formation of the thermalon appear in terms of a smooth field evolution in which thermal fluctuations are comparatively small or averaged-over? To answer this question, we can consider a field configuration $A_0$ very close to the thermalon, but which will re-collapse and return to (metastable) equilibrium. A small perturbation of $A_0$ would, then, lead to an expanding true vacuum bubble.  If $A_0$ is close to the thermalon, it will also be close to a ``bounce" configuration, and our prescription indicates that to obtain its formation history, we merely time-reverse the decay.  

\begin{figure*}[htb]
\includegraphics[width=15 cm]{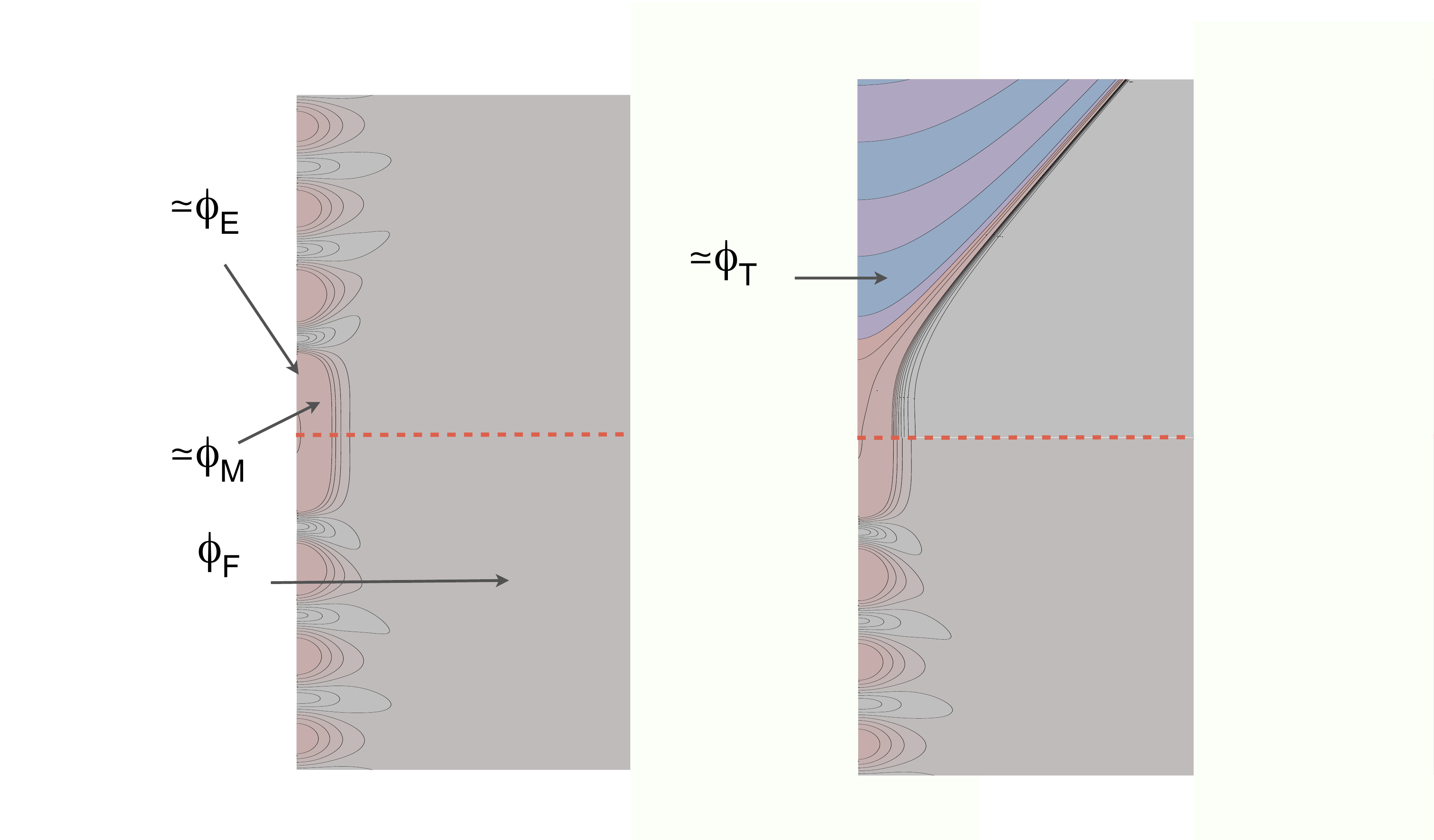}
\caption{
The ``thermalon", a static field configuration thermally fluctuated from a false vacuum.  The static configuration (computed numerically in flat space as described in~\cite{aguirre-2009-79})  is given on the spatial slice indicated by the dashed red line, and runs from field values somewhat past the potential barrier maximum value (see Fig.~\ref{fig-doublewell}) of $\phi_M$ near the lump's center, to the false-vacuum level $\phi_F$.  The configuration is here much smaller than the horizon scale (not shown). With a small perturbation the static configuration can decay back to the false vacuum (subsequent evolution in left panel) or to the true vacuum (right panel).  By our prescription, the fluctuation from the false-vacuum (metastable) equilibrium to the thermalon should be the time-reverse of the thermalon's decay to the false-vacuum: incoming scalar waves converge to set up oscillations that eventually reach the amplitude of the thermalon configuration, as depicted in the lower half of the left panel.  However, once the thermalon forms, it can also decay to the true vacuum, generating a full picture of purely thermal decay from the false vacuum (right panel.)
\label{fig-thermalon}
}
\end{figure*}

This is illustrated in Fig.~\ref{fig-thermalon}, which shows the numerical solution (using the method, potential, and numerical details of~\cite{aguirre-2009-79}) of an initial nearly-static configuration (on the horizontal dashed line).  This configuration has field values between $\phi_M$  and $\phi_E$ in the center, and decays to $\phi_F$ far away.  As it decays back to metastable equilibrium (upper half of left panel) it oscillates in the false vacuum, giving off ``waves" of scalar radiation that carry away the field's kinetic energy. Note that we are assuming that the temperature is low enough that thermal fluctuations in the field are small compared to the field excursions depicted in Fig.~\ref{fig-thermalon}.

By our prescription, this field configuration is formed by the somewhat counter-intuitive time-reverse of this: ingoing scalar waves emerge from the thermal bath and converge repeatedly on the same point, steadily exciting the field until it reaches the requisite amplitude and field configuration.  While this may seem unlikely, {\em all} significant excursions from equilibrium are, and our argument is simply that {\em given} that the static configuration is realized, it ``most probably" does so by this type of process. 

More precisely, we should consider the set of all possible decays to equilibrium {\em including} thermal fluctuations (which are not included in the solutions shown), and time-reverse this set of histories to obtain the set of formation histories expected for the thermalon configuration.  This set would be dominated by a particular history to the extent that thermal fluctuations are small compared to classical evolution.  

Once the thermalon forms, it also represents a significant decay channel to {\em true} equilibrium: in the static limit, the evolution becomes necessarily fluctuation-dominated, and the configuration can, with appreciable likelihood, decay into the true vacuum, as depicted in the right-hand-side of Fig.~\ref{fig-thermalon}.\footnote{Note that the time-asymmetry of the set of histories leading to and from the thermalon is not in contradiction with our prescription, because we have assumed that the history begins in a metastable equilbrium rather than the true equilibrium; the full set of possible histories leading up to the thermalon would include those originating near the true vacuum as well.}

Adding gravity, it is also possible to find static field configurations containing a pocket of the true vacuum. This instanton (see~\cite{Garriga:2004p3391}) is interpreted as describing the thermal activation of a pocket of true or false vacuum from the non-zero temperature of de Sitter radiation in the causal patch.~\footnote{In the ``thin wall" limit in which the domain wall between the true and false vacua can be considered an infinitesimally thin spherical membrane of constant tension, the thermalon and CDL decay (see Sec.~\ref{sec-cdl}) constitute, respectively, purely-thermal and purely-quantum limits of a family of processes parametrized by the Schwartzschild-de Sitter mass parameter outside the bubble, and described by an effective potential governing the 1-d dynamics of the domain wall's radius; see~\cite{Aguirre:2005nt}.} 
If gravity also follows our assumptions in Sec.~\ref{sec-prescription}, then the formation of such a configuration is just as for the non-gravitational case: scalar waves converge inward from the horizon to build up the thermalon field configuration, which can then evolve into an expanding or contracting pocket of true vacuum.

\subsection{Coleman-de~Luccia transitions}
\label{sec-cdl}

While the thermalon represents a purely thermal mechanism of decay from a thermal false vacuum, the most probable decay channel typically combines thermal activation with quantum tunneling.  This ``Coleman-de~Luccia" (CDL) process (for details see \cite{1977PhRvD..15.2929C,Coleman:1977py}, and \cite{Banks:2002nm,Brown:2007sd} for some recent analyses) is mediated by an O(4)-invariant Euclidean instanton that describes the classically-forbidden ``evolution" (the stationary trajectory of the Euclidean path integral) between the pre- and post-tunneling space and field configuration.  

\begin{figure*}[htb]
\includegraphics[width=12 cm]{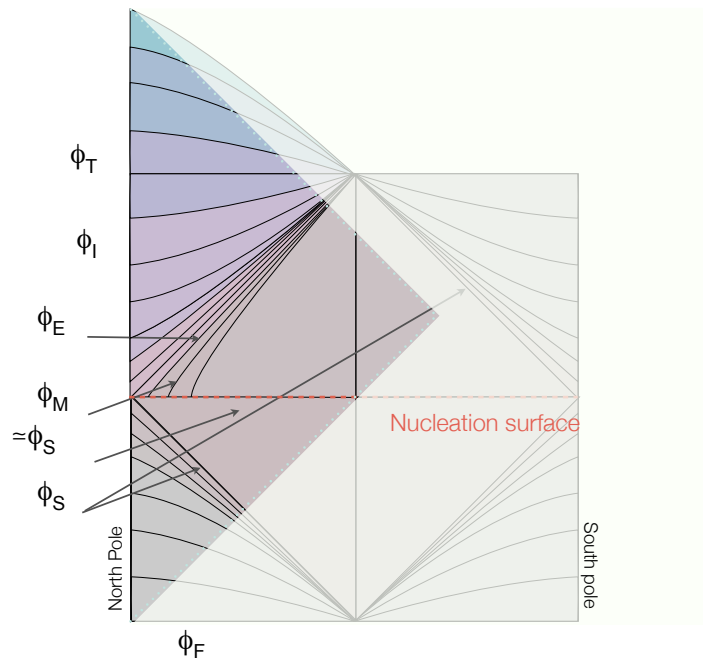}
\caption{
A depiction of thin-wall Coleman-de~Luccia tunneling.  The CDL instanton analytically continues to a spherically-symmetric space and field configuration shown just to the future of the ``nucleation surface" (red dashed line).  This evolves, in the ``north" pole (left side of diagram) into a true-vacuum bubble; around the ``south" pole the field relaxes back to the false vacuum.  The pre-tunneling configuration inside the causal patch (unshaded region) should be given by the south-pole {\em post} tunneling field configuration; it would then, by our prescription, arise as depicted in the bottom left quadrant of the diagram, with the field rolling up either (a) homogeneously, in the ``open" time slicing, or (b) inhomogeneously and inward, in time slicings covering the full spacetime.
  \label{fig-CDL}
}
\end{figure*}

This instanton is compact, and its analytic continuation yields an SO(3,1)-invariant spacetime with two ``bubbles", as depicted in Fig.~\ref{fig-CDL} (top half of diagram; see caption for details). In one bubble (at the ``north pole" center of spherical symmetry in the diagram), the field decays to the true vacuum; in the other (centered on the ``south pole") it relaxes back to the false vacuum. If the false vacuum has positive cosmological constant, the instanton endpoints do not coincide with either the true or false vacua, but rather between some value $\phi_S$ near the false vacuum (see Fig.~\ref{fig-doublewell}), and some value $\phi_E$ in the basin of attraction of the true vacuum (in ``open inflation" models, an inflating phase with $\phi \sim \phi_I$ takes place between the tunneled-to value and reheating near $\phi_T$.) 

There are two difficulties in interpreting the CDL instanton. The first has to do with its seemingly global nature: forming a true vacuum bubble at one pole seems to require an excitation of the false vacuum at the other, even though these two regions are forever out of causal contact.  Moreover, the instanton naturally describes a post-tunneling geometry of closed topology and small total volume; it is less clear how to interpret the tunneling process in a large expanding background including many dS Hubble volumes. The second difficulty is determining what exactly the pre-tunneling configuration should be, and how it arises. That the instanton endpoints do not reach the false vacuum is attributed to the ``thermal nature" of dS space, but is it possible to find a more complete description?

The interpretation advocated by Brown and Weinberg~\cite{Brown:2007sd} partially addresses both of these points. The first difficulty can be circumvented by restricting one's attention to the interior of the causal diamond, as indicated by the shading in Fig.~\ref{fig-CDL}, consistent with our picture. For the second, they propose that the pre-tunneling configuration is inhomogeneous, and given by the half of the instanton extending from the boundary of the dS horizon to the south pole of the instanton. It is this configuration that must be reached by a thermal fluctuation. 

How does such a configuration arise from empty dS? Our prescription indicates that this corresponds to the time-reverse of the relaxation to the false vacuum given by evolution of the south-pole end of the configuration (the shaded region of Fig.~\ref{fig-CDL}). An observer at the north pole witnessing this fluctuation would find themselves in an approximately open universe in which the field steadily roll up the hill to $\phi_s$. Widening the view to the entire causal diamond, the description would be of an inhomogenous configuration built up starting near the the horizon and propagating inward, eventually forming a configuration that is momentarily close to the SO(3,1) symmetric pre-tunneling configuration. With some luck, tunneling will then lead to an expanding true vacuum bubble. 

 Note that, because of the approximate SO(3,1) symmetry, there would be a family of geodesic observers related by boosts that would observe the same pre-tunneling field evolution (in the open slicing).  These observers would not, however, share the same causal diamond. So in specifying the ``system" we have also selected one such observer and hence a preferred frame.  It seems plausible that this frame is related to the frame picked out by the tunneling process (i.e. the surface across which the pre-and post-tunneling spacetimes are matched).

\subsection{Hawking-Moss}

\begin{figure*}[htb]
\includegraphics[width=12 cm]{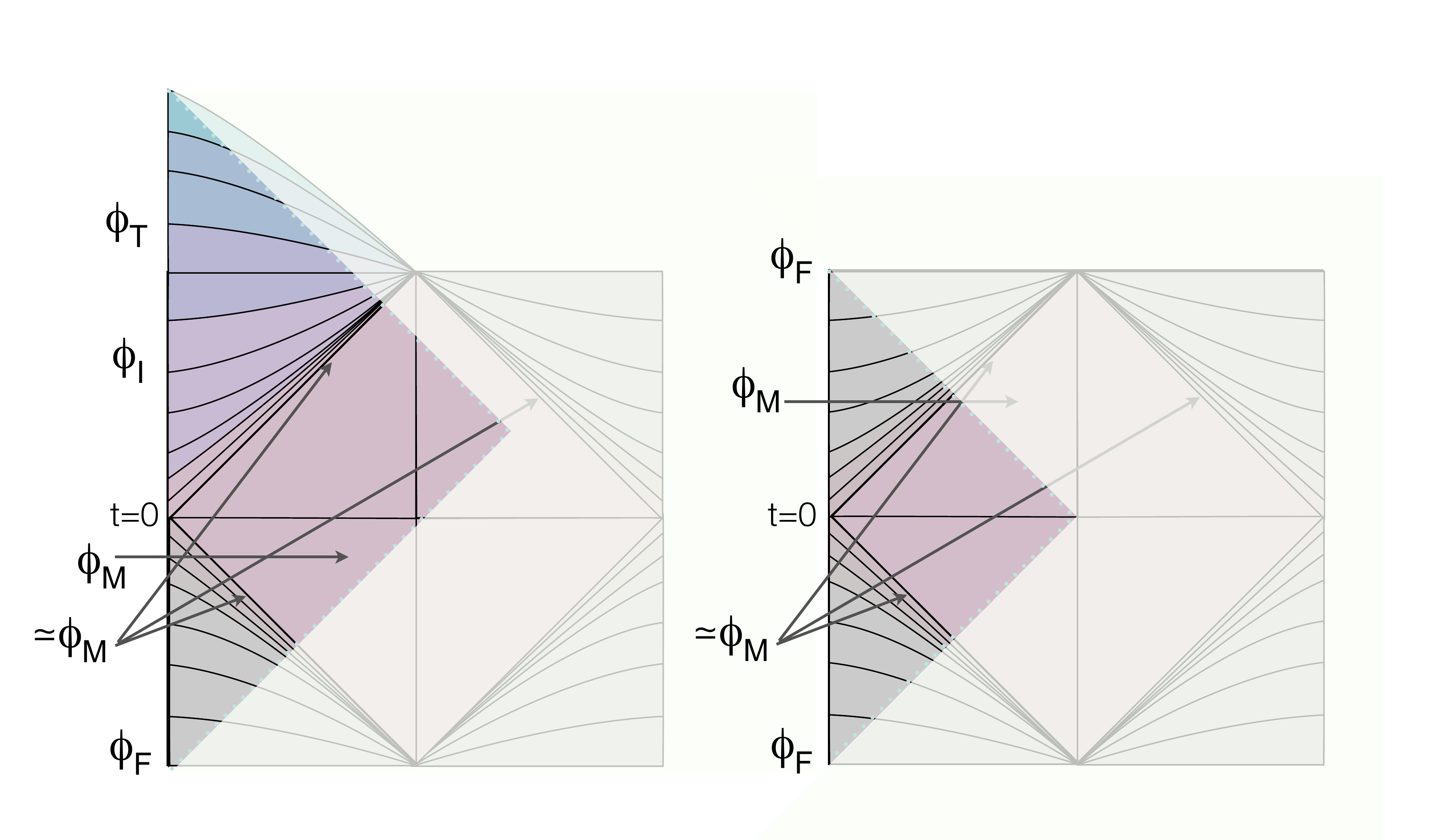}
\caption{
The Hawking-Moss process as a limit of a successful (left panel) and failed (right panel) CDL transition, in which the barrier potential becomes flat and the tunneling endpoints both approach the potential maximum. The result is the thermal fluctuation of a homogeneous configuration entirely composed of a thick uniform domain wall.  The field can then evolve into the true or false vacuum.  Note, however, that in the Hawking-Moss regime, the thermal fluctuations of the field cannot be ignored relative to the classical evolution, so this figure is simply one possible history among many of comparable probability.  
  \label{fig-HM}
}
\end{figure*}

The CDL solution exists only when the potential near $\phi_M$ is sufficiently curved relative to its height: $|V''(\phi)| \agt H^2$. If we consider a family of potentials with an increasingly broad or flat barrier, the instanton endpoints migrate increasingly near to the barrier maximum $\phi_M$, signifying that the thermal part of the process is increasing in importance and tunneling part decreasing. In the limit where the endpoints meet there is no CDL solution, but only the ``Hawking-Moss"~\cite{Hawking:1981fz} instanton, which is simply a 4-sphere covered with a homogeneous field $\phi=\phi_M$.  This instanton is generally interpreted as representing the purely thermal fluctuation of a horizon-sized region to $\phi_M$ (which is motivated by the matching of the Hawking-Moss tunneling probability to the probability obtained from the formalism of stochastic eternal inflation~\cite{Starobinsky:1986fx}); because this is unstable, the region will then evolve back to $\phi_F$, $\phi_T$, or a mixture of both.

How does this fluctuation occur? For a potential which admits a CDL instanton arbitrarily close to Hawking-Moss, the story will be the same as in the last section. This is depicted in Fig.~\ref{fig-HM}. Here, the pre-tunneling configuration is a nearly homogeneous horizon-sized region at $\phi=\phi_M$. In the absence of tunneling, the field relaxes back to the false vacuum equilibrium. The thermal fluctuation is the time-reverse of this relaxation process. From the thermally fluctuated field configuration, the vanishing tunnelling component of the evolution can create a bubble of true vacuum.  

For the true Hawking Moss instanton, this picture may be a bit too simplistic. The limit $|V''(\phi)| \alt H^2$ (where the CDL instanton vanishes) also coincides with the criterion for stochastic eternal inflation ({\it e.g.},~\cite{Batra:2006rz}). Thus the thermal fluctuations can never be considered small compared to the field's classical evolution, and there is no clearly preferred evolution from a non-equilibrium state back to equilibrium. A more accurate picture would probably be that the homogeneous field at $t=0$ fluctuates in some places toward $\phi_T$ and in some toward $\phi_F$, and breaks up into an inhomogeneous mix of true and false vacuum regions separated by inflating domain walls.  The evolution toward either vacuum might locally resemble the left- or right-side of \ref{fig-HM}, but the location of fastest descent (the origin in Fig.~\ref{fig-HM}) would be random. Likewise, the route from the false vacuum to $\phi_M$ would likely take significant excursions rather than rolling up steadily. In other words, although there might be a most likely history, the variance around it will be large. This can be made more precise in the context of stochastic eternal inflation, which we now describe.

\subsection{Stochastic eternal inflation}\label{sec:stochastic}

The standard treatment of stochastic eternal inflation involves a horizon-sized coarse graining, after which the evolution is described by either a Langevin equation governing the local behavior of the inflaton field or a Fokker-Planck equation governing the evolution of the probability distribution for the field (see {\it e.g.}~\cite{Linde:2005ht} and references therein). We shall focus here on the former. 

In the Langevin description, the field in each horizon volume is subject to a classical force from the potential and a stochastic force due to de Sitter fluctuations. The formalism is identical to that of Brownian motion in the presence of a one dimensional potential, and the time-reversal invariance of the microphysics underlying the quantum fluctuations is encoded in the statistical properties of the stochastic force. Because of this stochastic force, there is an ensemble of possible histories branching off from any initial condition, and the output of the Langevin analysis are the members of this ensemble. 

The Langevin analysis describes an unbiased coarse-graining of
time-reversible equations of motion for the perturbations in an
inflating universe. Therefore, the assumptions required to apply our
``prescription" for the picture of rare fluctuations are satisfied.
While there is no clear classical trajectory for the field in any particular horizon volume, there is a well defined average trajectory, an associated variance, and perhaps higher moments of the distribution as well. This makes precise how accurate our prescription might be in determining the history of a rare fluctuation.

Consider a scalar field with a potential that allows for stochastic eternal inflation, but also has a well-defined de Sitter equilibrium. For example, $V = \frac{1}{2}m^2 \phi^2 + V_0$, where $V_0$ is a constant positive vacuum energy at the equilibrium state $\phi = 0$. How might a fluctuation from equilibrium to a non-zero value of $\phi = \phi_0$ occur? Applying our prescription, we first ask how the field would evolve from $\phi= \phi_0$ back to equilibrium, and then time-reverse. The Langevin approach yields a full probability distribution of trajectories; the time-reverse of the mean trajectory is the most likely history for the fluctuation, with the uncertainty quantified by the variance of the distribution. 

If we take $m \ll m_p$ and $m \ll H_0$ with $H_0^2 =  V_0 / 3 m_p^2$, the equation of motion for the coarse-grained field is
\begin{equation}
3 H_0 \dot{\phi} = \frac{3 H_0^{5/2}}{2 \pi} n(t) - m^2 \phi(t),
\end{equation}
where $n(t)$ is a gaussian noise term of zero average and unit variance. The Langevin equation in this case admits an exact solution~\cite{Gratton:2005bi}:
\begin{equation}
\label{eq:solLang}
\phi (t) = \exp\left[-\frac{m^2}{3H_0} t \right]  
 \left( \phi_0  + \frac{H_0^{3/2}}{2 \pi} \int_0^{t} n(t') \exp\left[\frac{m^2}{3H_0} t' \right] dt' \right) . 
\end{equation}
Taking the ensemble average over trajectories, the average trajectory is given by
\begin{equation}\label{eq:mean}
\langle \phi (t) \rangle = \phi_0 \exp\left[-\frac{m^2}{3H_0} t \right] ,
\end{equation}
which is the slow-roll solution. The second moment of the distribution is
\begin{equation}
\langle \phi (t)  \phi (t') \rangle =  \exp\left[-\frac{m^2}{3H_0} \left( t +t' \right) \right]  \left( \phi_0^2 
+ \frac{3H_0^4}{8 \pi^2 m^2} \left[ \exp\left[\frac{2 m^2}{3H_0} {\rm min}\left( t,t' \right) \right] -1 \right]  \right)  .
\end{equation}
All other moments are consistent with being gaussian. The full probability distribution is given by:
\begin{equation}\label{eq:pdf}
\rho(\phi (t) | \phi_0) = \frac{1}{\sqrt{2 \sigma^2}} \exp \left[ - \frac{\left( \phi(t) - \langle \phi(t) \rangle \right)^2}{2 \sigma^2}\right],
\end{equation}
where
\begin{equation}\label{eq:variance}
\sigma^2  =   \frac{3H_0^4}{8 \pi^2 m^2} \left[ 1-\exp\left[-\frac{2 m^2}{3H_0} t  \right]  \right] .
\end{equation}

\begin{figure*}[htb]
\includegraphics[width=6 cm]{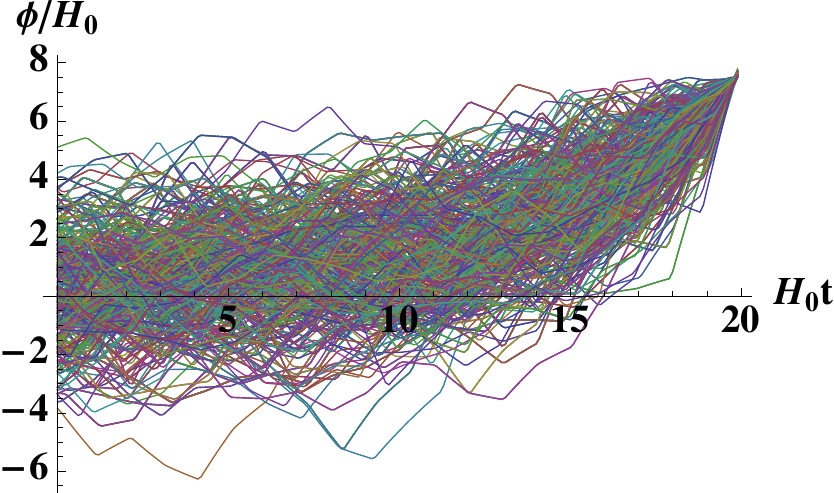}
\includegraphics[width=6 cm]{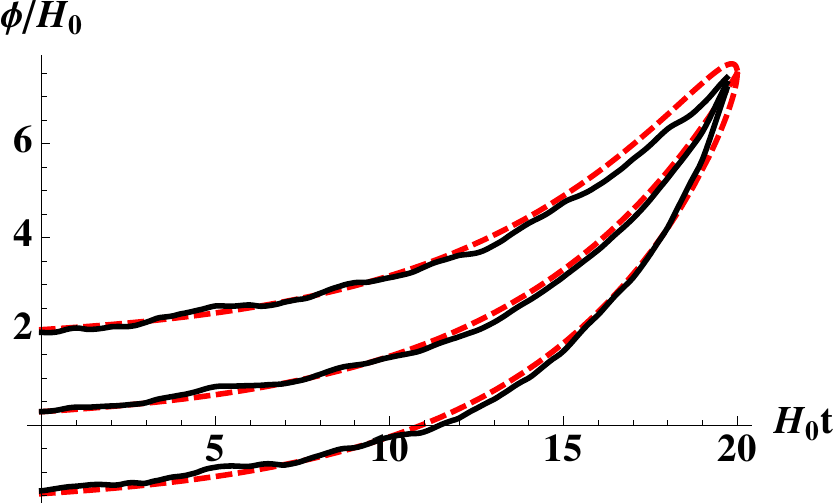}
\caption{Trajectories that reach $\phi_0 = 7.5 H_0$ drawn from the equilibrium probability distribution. Out of 500,000 total trajectories, only 629 reached this value in the allotted simulation time of $100 H_0^{-1}$. Each trajectory is time-translated to overlap. On the left we show all trajectories; 
on the right we show the average and variance of the trajectories shown on the left (solid black lines). These match the time-reverse of the mean and variance for trajectories going from $\phi_0$ back to equilibrium (dashed red lines).
\label{fig-tras}
}
\end{figure*}

By forming a large number of noise realizations, we numerically generated a set of 500,000 histories using Eq.~\ref{eq:solLang}. We use $m/H_0 =.7$ and run each trajectory for a time $t=100 H_0^{-1}$ (this is equivalent to looking at a single, long trajectory). After allowing time to reach equilibrium, we selected all histories that reach a pre-specified position $\phi_0$ (here, $\phi_0 = 7.5 H_0$). Focusing on a window of time around each fluctuation, and time-translating all trajectories so that they reach $\phi_0$ at the same time, we obtain the set of histories shown in Fig.~\ref{fig-tras} (left panel). The average and variance of the trajectories are shown in Fig.~\ref{fig-tras} (right panel). Both are in good agreement with the expectation that the probability distribution over histories going {\em from} equilibrium to $\phi$ is the time-reverse of the probability distribution from $\phi_0$ {\em to} equilibrium, which is given by Eq.~\ref{eq:pdf}. The time-reverse of the classical slow-roll solution will be a fairly good approximation to the history of the fluctuation. The error in this description is quantified by the variance, which in this example is necessarily quite large to maintain the validity of our approximate Langevin equation. 

\subsection{Black hole nucleation}
\label{sec-BH}

Black holes provide an excellent testing ground for exploring the thermodynamic properties of gravity. Indeed, the discoveries of Hawking radiation and black hole thermodynamics remain some of the best clues we have as to the nature of quantum gravity.  While black holes themselves are simple, there are many possible processes to study because black hole spacetimes can be diverse. Black holes can form in various backgrounds, and because time-reversal of a black hole yields a white hole, discussion of these entities is necessary as well.

Considering first flat space at finite temperature $T$, it was shown by Gross, Perry, and Yaffe~\cite{Gross:1982cv} that there is a finite probability to produce black holes from the thermal bath. The post-fluctuation configuration they proposed was a black hole of mass $M = 1 / 8 \pi G T$. Although such a black hole emits Hawking radiation of temperature $T$ and is in thermal equilibrium with the radiation bath, the equilibrium is unstable. So, the black hole is on the cusp of either growing indefinitely by absorbing radiation (and progressively decreasing its temperature), or shrinking to zero size by emitting Hawking radiation at progressively higher temperatures. This is very reminiscent of the thermalon discussed above, and in analogy with that case we would expect that this nucleation processes can be imagined as the limit of the formation of smaller black holes, which return to equilibrium by emitting Hawking radiation.\footnote{Because of the two different temperatures there is, however, no regular instanton describing the formation of these smaller black holes.}

How would such black holes form? Here our prescription is less clear than in previous cases: it indicates that we should ask how the time-reverse of a black hole -- a {\em white} hole -- would return to equilibrium, then time-reverse this history.  Unfortunately, this is unhelpful: all manner of material could in principle be expelled from a white hole and then decay to equilibrium, and we therefore get little insight into the {\em generic} formation history of a black hole.  

\begin{figure*}[htb]
\includegraphics[width=12 cm]{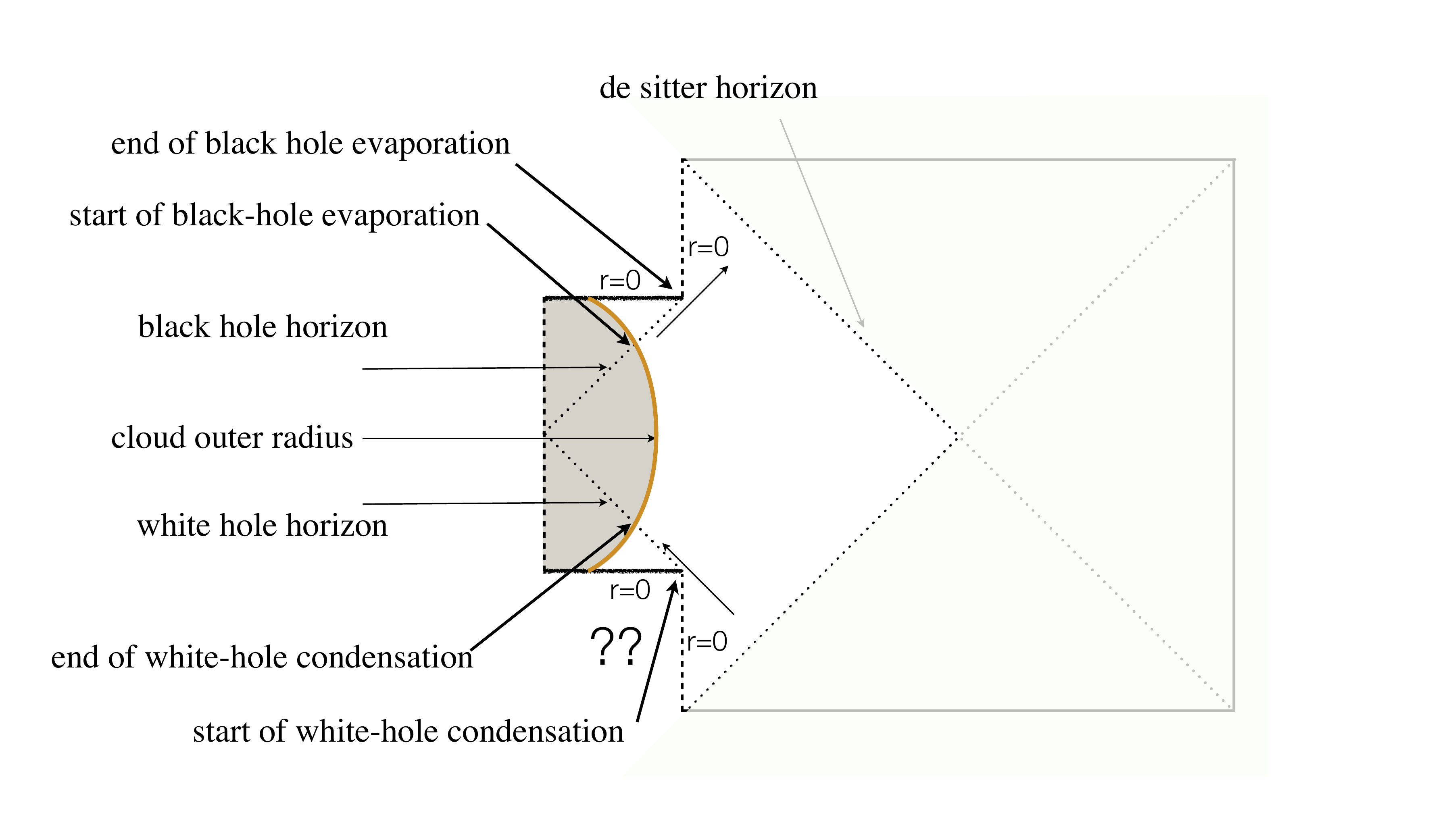}
\caption{
Depiction of black hole formation via the nucleation of a ``bounce" configuration of a ball of dust at rest, which collapses to form a black hole and (by the time symmetry of such fluctuations in our prescription) that was emanated from a white hole. In this non-unitary account of black hole formation and evaporation, we cannot apply our prescription to determine how the white hole itself was formed.
  \label{fig-BHsing}
}
\end{figure*}

We might, however, obtain some insight by examining a special class of histories in which there is a ``bounce" configuration that evolved {\em into} a black hole in the future. The time-symmetry then specifies which configuration had to be spewed out of a white hole to the past. Consider, for example, an idealized spherical ``dust ball" at rest.  This is described by the Oppenheimer-Snyder~\cite{Oppenheimer:1939ue} solution.  Now the ``classic" information-destroying account of this solution would be that the ball crosses the event horizon and eventually merges with a spacelike black-hole singularity. The black hole then evaporates via Hawking radiation. In this account, there is an obstruction to applying our prescription to determine the full history of the fluctuation. While we can specify the classical evolution of the dust shell out of the white hole, because the evaporation process is manifestly non-unitary, we cannot apply our prescription to determine how the white hole itself formed from equilibrium. See Fig.~\ref{fig-BHsing}. 

\begin{figure*}[htb]
\includegraphics[width=12 cm]{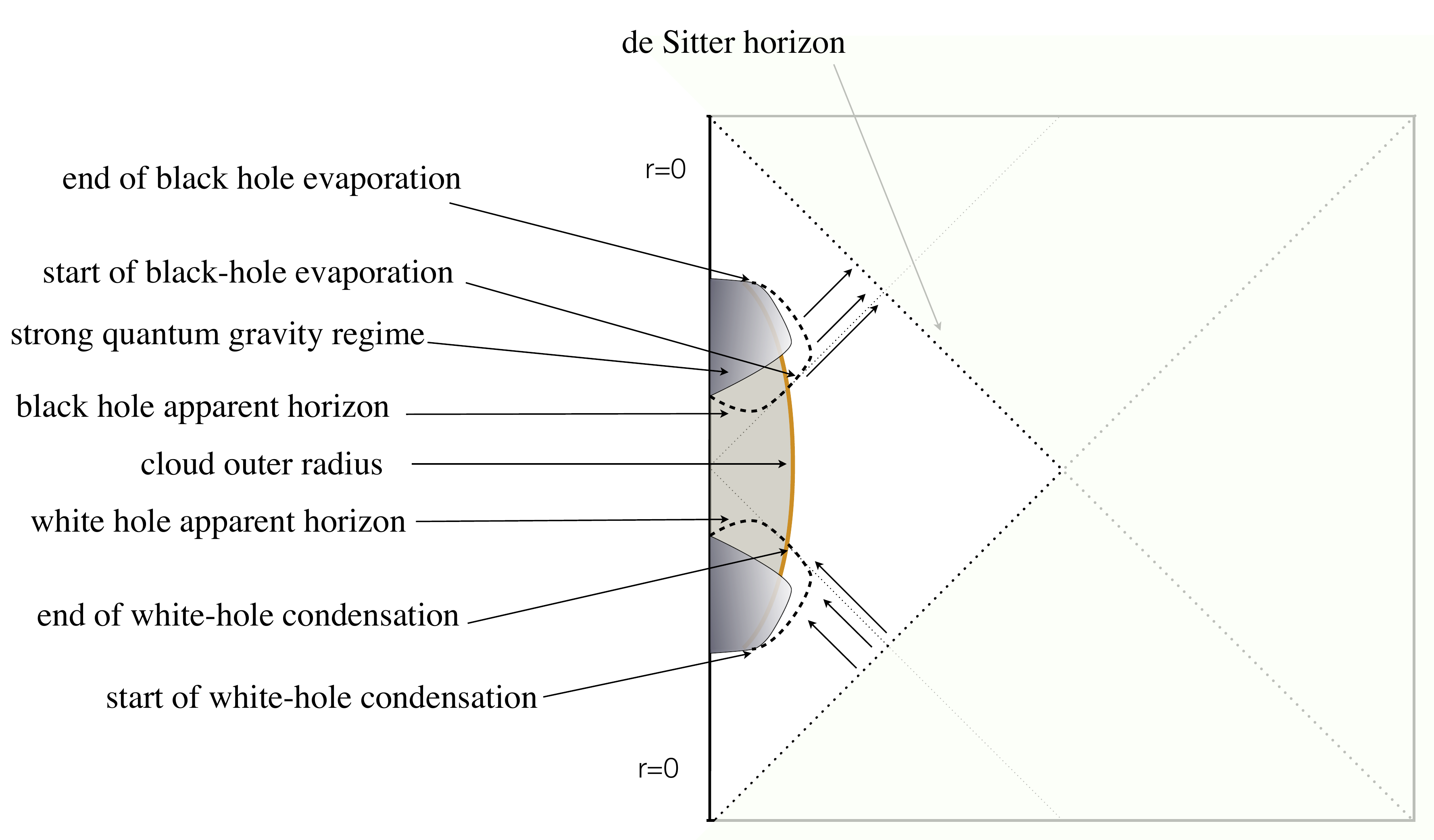}
\caption{
Depiction of black hole formation via the nucleation of a ``bounce" configuration of a ball of dust at rest, which collapses to a region in which classical gravity breaks down, but quantum gravity is presumed unitary. For a long period the apparent horizon is spherical and of the Schwarzschild radius; Hawking radiation forms near this horizon and escapes to infinity. Prior to the bounce, is a similar quantum-gravity region that accretes thermal radiation from far away, and eventually ejects an expanding dustball.
  \label{fig-BHunit}
}
\end{figure*}

If, however, black hole evaporation is unitary (as suggested by AdS/CFT and similar arguments), we can apply our prescription, and try to draw approximate conformal diagrams for fluctuations that ``shade out" some region where quantum effects on the spacetime become very large (see {\it e.g.},~\cite{Ashtekhar:2005p3384,Hayward:2006p3382,Horowitz:2009p3383,Hossenfelder:2010p3381}). This suggests a picture (see Fig.~\ref{fig-BHunit}) in which the ball crosses only an apparent horizon, which for some long time is of approximately the Schwarzschild radius. Near this radius outgoing Hawking radiation is generated that carries positive energy away, and negative-energy radiation propagates inward to shrink the enclosed mass. The dustball {\em formation} process would similarly involve both positive and negative-energy fluxes, but in which negative-energy radiation flows outward from the congealing center, and positive energy radiation converges from far away. We thus see that cosmic censorship (vaguely defined) is violated, but only with concomitant violation of the Second Law.  Note that even if we replace the dustball by a more generic configuration, the black hole always decays in the same way, via Hawking radiation.  Therefore, the formation of a white hole of mass $M$ {\em generically} occurs when a burst of ingoing radiation forms a core, followed by Hawking ``condensation": spherically-symmetric thermal radiation converges upon the seed to slowly build up the mass until it reaches $M$.

Let us now discuss black (and white) hole formation in various contexts. Our first example was thermal flat space at temperature $T$. From this (quasi) equilibrium we would expect to nucleate many small-mass time-asymmetric black holes (formed by an unknown generic process but decaying in a well-defined way), time-asymmetric white-holes (formed by a known generic process of Hawking condensation but decaying in an unknown way), and time-symmetric black-hole/white-hole pairs (growing and decaying via Hawking radiation). Eventually, however, in a given region a black hole of mass $M \agt 1 / 8 \pi G T$ will nucleate, then continue to grow indefinitely because the original spacetime was only a quasi-equilibrium state. 

\begin{figure*}[htb]
\includegraphics[width=14 cm]{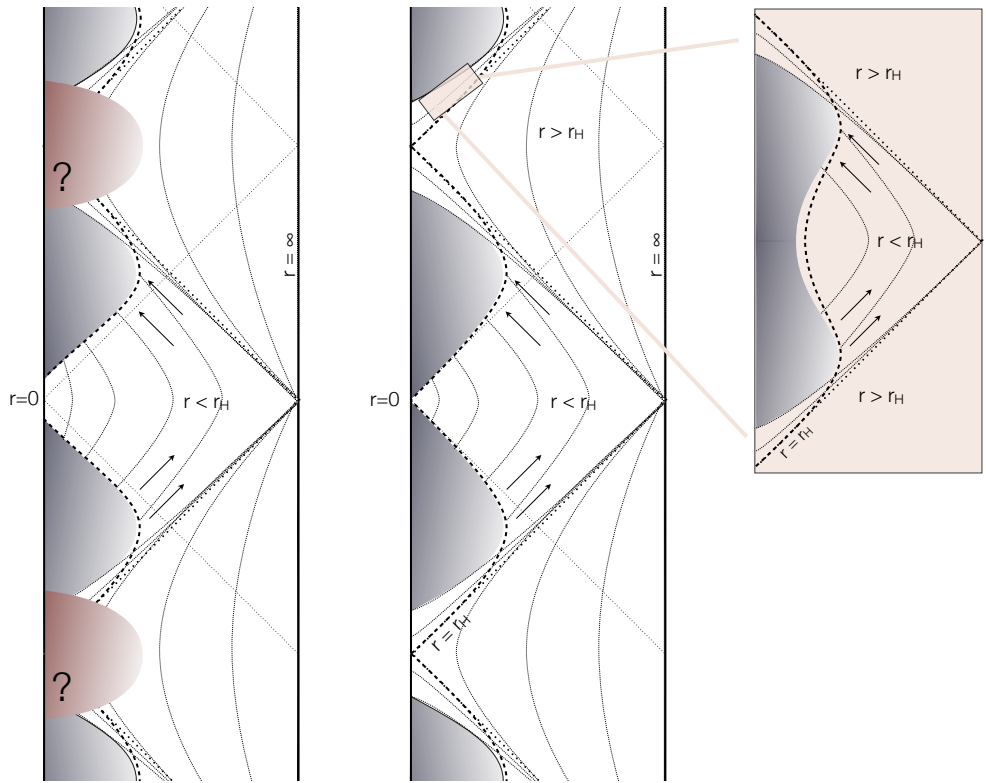}
\caption{
The conformal diagram for black-and-white holes in AdS. As in Fig.~\ref{fig-BHunit}, thick dashed lines are apparent horizons; thick dotted lines are lines of constant radius $r_H$, the horizon size of the black/white hole during its most long-lived phase. Thin dashed lines represent constant radii in (quasi)-static coordinates.  The {\em left panel} depicts AdS at low total energy; in this background, an unknown process (lower shaded `?' region) creates a high-density region including a trapped surface. This exists as black-hole-like object for an evaporation time, then decays into Hawking radiation. Some time later, a white hole may form via spherically-symmetric thermal ``Hawking condensation".  It then exists as a white hole for some time before spewing out an unknown result (upper shaded `?' region) that might recollapse into a black hole (as in the `dust ball' case above), or disperse and thermalize. The {\em right panel} depicts a large black hole in equilibrium with high-temperature AdS, which may fluctuate into thermal AdS. For an extremely long time, an observer at fixed radius sees a black-and-white hole solution.  Eventually, the black hole evaporates (while decreasing the system's total entropy) until it completely disappears for some time until a new `seed' forms, from which a white hole condenses while increasing entropy. (This process can be probed by an observer at small radius, but the observer must be very careful to enter at the right time, when the border of the strong-field region is timelike and can be avoided.) This white hole, and an accompanying black hole, then exist for a very long time again.  Between such evaporation events will occur exponentially more {\em partial} evaporation events, depicted in the zoomed box, in which the apparent horizon shrinks for some time, then re-expands to its equilibrium value.
  \label{fig-BHeternal}
}
\end{figure*}

The AdS/CFT correspondence gives strong support for the validity of our assumptions. Therefore, perhaps the most rigorous application of our prescription arises when discussing black holes and radiation in AdS. In asymptotically AdS space, the equilibrium configuration depends, in the microcanonical ensemble, on the total energy of the black hole and radiation~\cite{Hawking:1982dh}. At energies much lower than $m_p^2 |\Lambda|^{-1/2}$, the equilibrium configuration is a radiation bath with no black hole. Fluctuations in this case would lead to the formation of small black (or white) holes, which then evaporate back into equilibrium. This picture, illustrated in Fig.~\ref{fig-BHeternal} (left diagram), is very similar to the case discussed above for the formation of black/white holes in thermal flat space. 

At energies much larger than $m_p^2 |\Lambda|^{-1/2}$, the equilibrium configuration is described a black hole of radius larger than $\ell_{AdS}$ along with a radiation bath with a temperature matching that of the black hole. From the standpoint of this paper, however, two important notes should be made.  First, because this is a genuine equilibrium state, it should be time-symmetric and thus a ``black-and-white" hole; indeed this is manifest in the standard maximally extended conformal diagram of S-AdS (see, {\it e.g.}~\cite{Maldacena:2001kr}). 
Second, even an equilibrium system is not truly ``eternal": while it is entropically favorable for there to be a black hole, fluctuations will at least temporarily cause it to partially or wholly evaporate. The stable black hole configuration would then re-form from the thermal bath. Thus while {\em most} of the time there would be a single black hole and a single white hole available for viewing, a very patient eternal observer would witness repeated black and white hole formation and evaporation processes. The full history of this process as obtained through our prescription is explained in Fig.~\ref{fig-BHeternal} (middle diagram).

Put together, these views address a dispute between Hawking and Penrose spanning several chapters in~\cite{Hawking:2010p3661}. Hawking argues that because equilibrium should be time-symmetric, and black holes nucleate from thermal equilibrium, that they too, must be time-symmetric and hence also white holes; Penrose argues that most black holes are manifestly not time-reversible.  Our arguments suggest both are partially right: both time-symmetric and time-asymmetric black/white holes can fluctuate from equilibrium; the former would contain a ``bounce" configuration as in our dust-ball model.  Yet it is also true that the equilibrium state must be time-symmetric; this symmetry is manifested in the {\em statistical} description of all processes (including black and white hole formation), not necessarily in each individual object. 

In addition to providing a firm footing on which we can apply our prescription, the AdS/CFT correspondence can possibly be used to explore the properties of fluctuations in much greater detail. For example, a mapping between fluctuations in the thermal CFT and fluctuations in the bulk would yield a distribution over the possible black hole formation and evaporation histories. This distribution would in principle provide the generic formation and evaporation histories that are missing from the purely bulk  description presented above, yielding an average history and its associated variance. This restores the predictability lost when considering the possible configurations that might be spawned from a classical white hole.

Finally, we turn to de Sitter space. Here, for a given vacuum energy there is a maximum radius ``Nariai" black hole that can fit inside the dS horizon; all smaller uncharged black holes are unstable to evaporation. In dS, the quasi-equilibrium configuration that can be nucleated is a black hole slightly smaller than the Nariai black hole~\cite{Ginsparg:1982rs,Bousso:1995cc}. As for the thermalon or the flat-space black hole, we could imagine a perturbation of this bounce state either evaporating, or growing to fill the horizon volume. The formation process would follow as above.

Black holes much smaller than the Nariai black hole are far from bounce states, but presumably will form from the thermal bath. It is possible to estimate the rate for this to occur by calculating the difference in entropy between the initial and final states, as measured by the total area of the black hole and cosmological event horizons. Placing a black hole in dS space always causes the total area of the horizon to decrease, lending credence to the fact that empty dS is indeed an equilibrium state (the degrees of freedom making up the black hole can then perhaps be thought of as being ``borrowed" from the cosmological horizon~\cite{Susskind:2011ap}).

\subsection{The ``island universe" scenario}

	If essentially anything can be fluctuated out of empty dS, one may ask whether our entire observable universe may have, at some point in the past, fluctuated directly out of the vacuum.  This scenario has been proposed by Dutta \& Vachaspati~\cite{Dutta:2005p3435} and dubbed the ``Island universe" model (see also a related scenario by Piao~\cite{Piao:2008p3454}).  These authors envision the fluctuation as a very brief period in which quantum fluctuations that violate the weak energy condition (WEC) shrink the cosmological horizon while generating a hot thermal bath of radiation and matter, which would subsequently evolve as a standard hot big-bang cosmology.  Modeling the process as being driven by ``phantom" energy with equation of state $w < -1$ leads to a postulated period of super-exponential expansion.
	
	Following our arguments, the process would appear rather different.  In empty dS, suppose that we wait for the fluctuation of a configuration $A_0$ that is (a) nearly-homeogeneous, (b) expanding with scale factor $R(t)$, (c) consists of some specifiable combination of particles and fields, and (d) is thermal at some temperature $T$.  By suitably tailoring these specifications, and choosing $T \agt 10^{10}\,K$, we could obtain a region that would evolve like our observed universe.  As a simple model of such a process, we might imagine awaiting the fluctuation of a finite ball of radius $R_b$ and flat geometry centered at the origin, with Schwarzschild-de Sitter (SdS) geometry outside of the ball; the properties of such a cosmology have been worked out in detail by Adler et al.\cite{Adler:2p3526}. Demanding that the past lightcone of a point at our given matter density ($\rho_0$) remain inside the ball yields a minimal comoving ball radius $r_b$ and mass $m_b \equiv (4\pi/3)\rho_0 (R(0) r_b)^3$. For a ball of pure dust, the FRW interior can be smoothly matched to the SdS exterior of mass $m_b$~\cite{Adler:2p3526}.  
	
	If the fluctuated configuration $A_0$ is long before the time of density $\rho_0$, then $m_b$ exceeds the Nariai mass; it will be more convenient and clear to assume otherwise, so that the ball begins within the cosmological horizon.  We may now ask how $A_0$ most likely forms.  Because $A_0$ is not a ``bounce" configuration, we do not expect the entire history to be time-symmetric. To find it, we time-reverse $A_0$ to obtain a contracting, nearly-homogeneous ball.  As the ball contracts, the inhomogeneities grow (as $R^{-2}$ in radiation domination or $R^{-3/2}$ for matter domination; {\it e.g.}~\cite{Peacock:1999p3566}) and it becomes rather messy.  But because at this time the matter in the ball obeys the strong energy condition, by standard singularity theorems ({\it e.g.}~\cite{1975lsss.book.....H}),\footnote{Or, for that matter, the Tolman-Oppenheimer-Volkoff limit on stellar mass.} the region inevitably collapses into a singularity. 
	
	  In this case it seems fairly clear that a black hole forms, since in the case of strict homogeneity this is just the Oppenheimer-Snyder collapse discussed in Sec.~\ref{sec-BH}, and the inhomogeneities presumably just complicate this formation. Assuming negligible radiation from the ball during collapse, this black hole has mass $m_b$, and eventually evaporates to leave empty dS.
	  
\begin{figure*}[htb]
\includegraphics[width=12 cm]{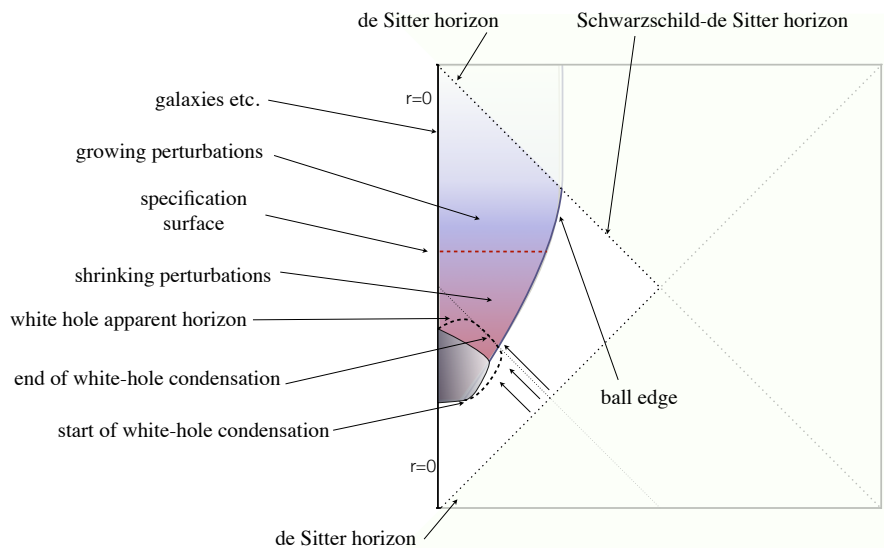}
\caption{
 Interpretation of the ``island universe" scenario, in which a large, expanding, radiation-dominated region fluctuates out of dS.  We model this as a finite ball, within which the FRW metric holds, and outside of which the SdS metric holds.  (The matching is straightforward for a dust interior, but more subtle with radiation inside.) For ease of analysis we assume here that the Schwarzschild mass $M$ is less than the Nariai mass (although this probably would not be observationally viable).  In this scenario we specify that we will wait until a nearly-homogeneous, expanding, radiation-dominated configuration appears (at a time depicted by the red dashed ``specification surface" curve.)  This region evolves like our observed universe, eventually becoming $\Lambda$-dominated and returning to dS.  To ascertain its most probable formation history, we time-reverse the macrostate to find a slightly inhomogeneous, contracting, radiation-dominated ball.  This collapses into a black hole that eventually evaporates.  Time reversing this in turn, a typical formation history is the condensation of a large white hole, from which springs an ultra-dense ball of radiation.  (We assume the ``unitary" version of WH nucleation discussed above).
 \label{fig-island}
}
\end{figure*}

	Time-reversing this history, the most probable formation mechanism is to condense a large white hole of mass $m_b$; this then expels a highly inhomogeneous ball that expands and smooths out until it matches the specified state $A_0$; subsequently the fluctuations begin to evolve as usual.  Eventually, the ball approaches empty dS again. This whole history is depicted in Fig.~\ref{fig-island}. 
	
	Comparing to the description of Dutta and Vachaspati~\cite{Dutta:2005p3435}, there are several key differences. The whole process is quite gradual, rather than sudden; there is no period of super-exponential expansion; and the form that the WEC violation takes is well defined, as discussed in Sec.~\ref{sec-BH}. This perspective on the scenario also makes manifest certain grave problems with it.\footnote{Another version by Piao~\cite{Piao:2008p3454} assumes that a fluctuation occurs in going from a {\em false} vacuum into an {\em excited} state of the true vacuum (rather than to inflation, which is presumed inoperative).  
This scenario is in some tension with the view we have advocated, because such a system (when time reversed) would tend to decay to the true, rather than false vacuum, and hence such a system would tend to arise from the true vacuum. (This is an instance of the general fact that without further specification, a given system is more likely to arise from a high-entropy precursor than a lower-entropy one.)  However the same concern could be leveled against open inflation, and the details are tied up here in the questions of measures, etc. in eternal inflation, to be discussed below.}

  First, suppose we merely want to wait for a configuration that includes observers (or planets or stars of a galaxy or some other proxy).  Then it is exponentially more likely  to fluctuate a low-mass ball (if the rate is set by the exponential of the difference in horizon entropies of the initial and final configurations) that is just sufficient to evolve into (say) a galaxy, rather than something of our current horizon size. Second, in order to explain the local universe {\em now}, precursor-balls at progressively earlier times must have progressively less entropy -- all require the same decrement in cosmological horizon entropy, but once specified, the ball's entropy must grow as it expands.  Thus given a specification of the presently observed state, it is also exponentially more likely to fluctuate our state $A_0$ very recently than far in the past.  Third, if we wish to fluctuate a configuration that is both early, and also covers our entire past lightcone -- as per the ``Island universe" scenario, the exterior solution is as for a SdS black hole with mass above the Nariai mass.  This has a global past singularity, so when we time-reverse the desired configuration, it just collapses into a global big crunch, and it is difficult to see how this can return to dS -- and thus how it could have emerged from dS.

\subsection{The Lee-Weinberg/``Recycling" process}

A quite different mechanism by which a positive true vacuum might ``restart" high-energy inflation and cosmological evolution is given by an interpretation of the CDL instanton, pointed out by Lee \& Weinberg~\cite{1987PhRvD..36.1088L} and incorporated into a ``recycling" cosmology by Garriga \& Vilenkin~\cite{Garriga:1997ef}. In this picture the CDL instanton is seen as interpolating between the true vacuum and a (large) false vacuum bubble. A puzzling property of this matching is that if it is done -- as for the CDL process -- along a surface of time symmetry of the instanton, then more than a full horizon volume of true vacuum must be replaced with a relatively tiny false-vacuum Hubble volume (see, {\it e.g.},~\cite{Garriga:1997ef,Aguirre:2005nt}). This can be avoided~\cite{Garriga:1997ef} by matching across a surface of constant flat-slicing time so that the replaced true-vacuum volume is equal to the false-vacuum volume removed; but the region then contains many, many, false-vacuum Hubble volumes.

\begin{figure*}[htb]
\includegraphics[width=12 cm]{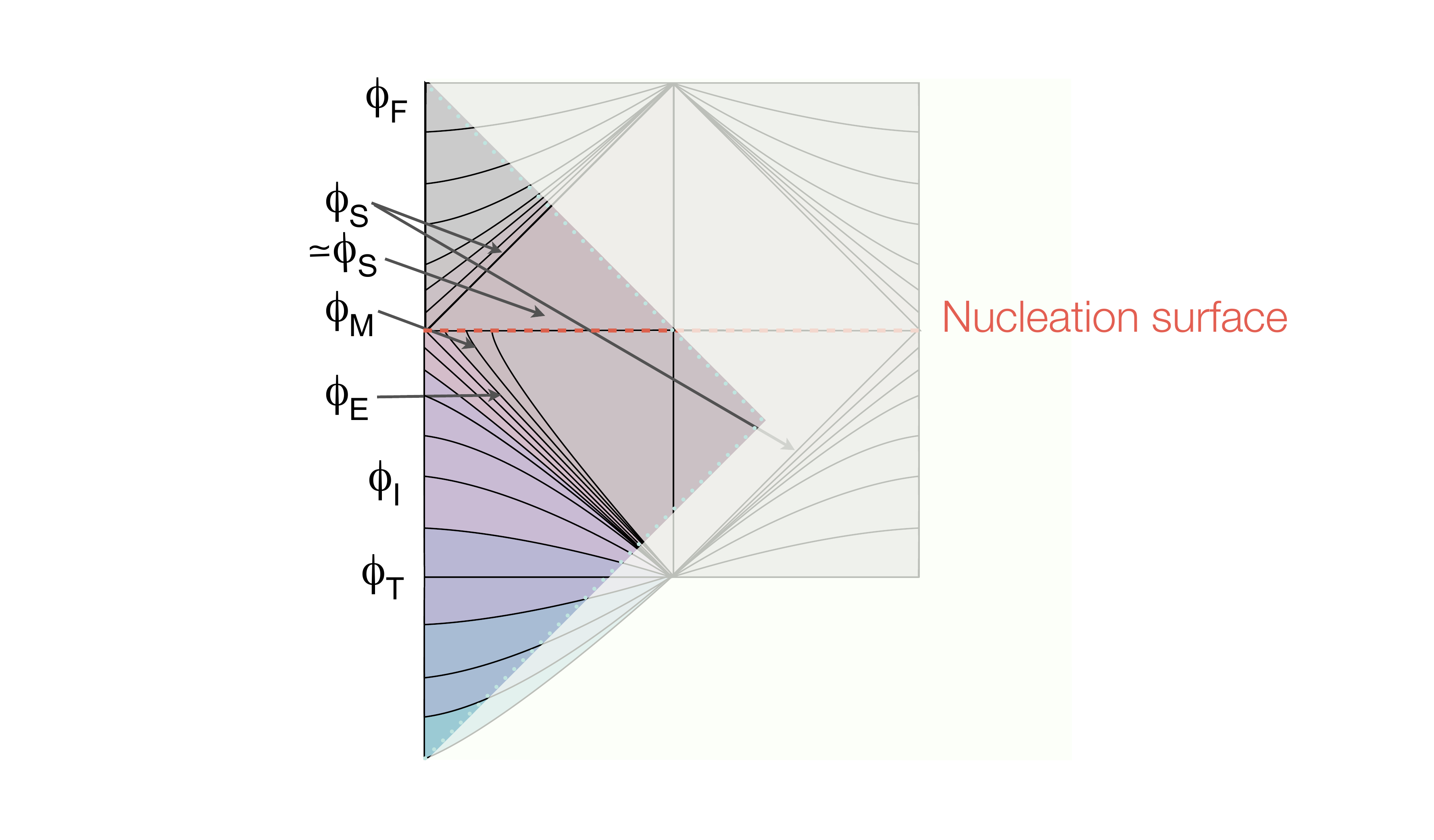}
\caption{
A view of the Lee-Weinberg or recycling process as a full time-reverse of CDL bubble formation.  Starting from empty dS, after an unspeakably long time nonlinear arrangements of matter congeal, and standard cosmological evolution runs in time-reverse through the vacuum, matter, and radiation-dominated epochs (with shrinking density perturbations).  The radiation coherently excites the inflaton field, which is impelled uphill and given anti-friction from the cosmic contraction, so that it slow-rolls up to form a small bubble of true vacuum, which can ``denucleate" via tunneling to form a horizon volume of pure false vacuum. 
   \label{fig-LW}
}
\end{figure*}

The view suggested in the preceding sections is that the formation of a false vacuum bubble includes a significant thermal component (as the instanton endpoint is far from the true vacuum, if sufficient slow-roll inflation is to occur inside the bubble). If we take this view, and confine consideration to the causal diamond of an observer, we obtain a picture which is almost precisely CDL, but time-reversed (see Fig.~\ref{fig-LW}). The recycling process in this interpretation would, then, look like a time-reverse cosmological evolution -- vacuum domination through matter then radiation domination, until the inflaton field rolls steadily uphill (with a steadily decreasing Hubble radius), finally coming to rest as a small bubble. This configuration could then presumably ``denucleate".\footnote{Our prescription does not really address the denucleation process, but we shall assume that there is nothing fundamentally different in tunneling through the barrier in one direction or the other. This is motivated by the interpretation of the CDL instanton as describing either the nucleation of a true or a false vacuum bubble.}

In considering this history it must be kept in mind that essentially {\em any other possible} set of events contained within the true-vacuum horizon are exponentially more probable. The outlined history is likely only relative to {\em other} ways of creating a horizon volume of false vacuum, given our other assumptions. This again highlights the improbability with which inflation can be initiated from a non-inflating phase, and is close kin to the ``Boltzmann Brain" paradox (those objects being discussed in the next section)  and to the difficulties for the ``Island Universe" discussed in the previous section. 

In models of eternal inflation, there are a number of proposed ways in which this problem can be avoided.  Most of these start by assuming (at various levels of rigor) that there {\em is} no true equilibrium state. Some explanations of this include: assuming there is a fixed but infinite set of states ({\it e.g.}~\cite{Carroll:2004pn,Carroll:2008yd}), allowing for the effective creation of new states via the cosmic expansion ({\it e.g.}~\cite{Kofman:2002cj}), 
or assuming the system generally does not exist long enough to really attain equilibrium ({\it e.g.},~\cite{Page:2006nt,DeSimone:2008if,Bousso:2008hz}).

However, the lack of a true equilibrium does not necessarily suffice to solve the problem. This is because many measures in eternal inflation predict that probabilities will be dominated by the decay products of the longest-lived, generally very high entropy, vacua ({\it e.g.},~\cite{Simone:2008p3104}). Thus to avoid domination by ``freak" histories, further explanation is required. Three proposed possibilities are:
\begin{enumerate}

\item A mechanism, such as that of~\cite{Farhi:1989yr,Fischler:1990pk}, by which a region of inflating false-vacuum could be created with much higher probability than via the Lee-Weinberg process; see, {\it e.g.},\cite{Albrecht:2004ke,Carroll:2004pn,Aguirre:2005nt,Albrecht:2009vr}; the accompanying notion is that inflation is cheap.

\item The measure strongly rewards paths that descend from high-energy, rendering recycling irrelevant ({\it e.g.},~\cite{Linde:1980tt,Linde:1993xx,Linde:2006nw}).  This accords with the notion that although it might be ``expensive" for inflation to start, once it does so, it creates infinitely many instantiations of any given $A_0$, rendering each one low- or zero-cost.

\item If the slowest-decaying vacuum cannot support any observers, then the system must transit through a false vacuum, then inflationary phase, before it can create observers, giving them a ``normal" formation history~\cite{Simone:2008p3104}.  Here the notion is that inflation is expensive, and may or may not pay back infinite return, but at any rate is the only game in town.\footnote{Note that this mechanism would not save a scenario with a true equilibrium and satisfying our other conditions: the ``anthropic" vacuum would simply be a metastable equiliubrium, and an excitation of this state would be much more likely to arise from that equiliubrium than directly from the true one.}

\end{enumerate}

Each of these ideas is in some tension with our treatment because they rely on a wider context than our ``system" of the causal diamond of a single observer.  The first explanation, for example, probably forbids any observer from entering the new inflating region (because the nucleated bubble is separated from the parent spacetime by a wormhole; this is the Farhi-Guth-Guven process~\cite{Farhi:1989yr}). This greatly increasing the difficulty of defining the ``system", or of defining probabilities for histories~\cite{Aguirre:2006ak}. The second and third explanations involve discussing thermal fluctuations (the precursors to false vacuum bubbles and freak observers) on time and distance scales inaccessible to a single observer. In the global picture the pre-tunneling field configuration for the Lee-Weinberg process includes a region of false vacuum much larger than the size of the horizon, and by causality it is not clear how this could evolve classically back to the true vacuum (and thus our prescription would appear useless).\footnote{One possibility is that a very rare percolation event occurs, whereby true vacuum bubbles conspire to form quickly enough to completely destroy the false vacuum. Time reversing this, some variant of our prescription would be that the true vacuum spontaneously fragments into collapsing bubbles, which then de-nucleate into the false vacuum. If this is the correct picture, one would then have to re-evaluate probabilities to include the rarity of such a percolation event.}  Moreover, both scenarios seem to solve the problem by giving ``credit" to the volume produced by inflation on super-horizon scales, but don't charge a price related to all the volume destroyed by deflation.

\subsection{Boltzmann Brains}

As a final application, we turn to one of the most bizarre objects that might spring from equilibrium: a fully-formed, thinking, sentient brain~\cite{Dyson:2002pf,Albrecht:2004ke,Linde:2006nw}. This possibility is of potential interest due to the argument that the most probable fluctuation necessary to explain a given set of observed data would just be a single observer, duped into believing that that such data had been observed.  This requires, so the argument goes, just a single brain and enough surroundings to support its operations for a very brief period of time.  We will not discuss the details of this argument here, but rather address the question of how such a brain would actually arise from equilibrium.

Here, the state $A_0$ that defines the boundary condition into which the system fluctuates would be a functioning brain, along with say a few tens of meters of life-support system so that it can process information normally for a minute or so before something goes awry.  Following this minute, the brain would ``die", and return to equilibrium.  This would take an extremely long time: parts of the life-support structure might evaporate quickly, but soon the brain and the remainder of its support would freeze, desiccate, and sit for a vast time until rather unlikely processes such as tunneling of its atoms and/or proton decay complete its disintegration.

To generate a formation history, we must first time-reverse the macrostate $A_0$ into $\bar A_0$ and assess how it would evolve.  This raises a thorny question: how well does the macrostate $A_0$ have to be specified to ensure a functioning brain? (For some discussion see {\it e.g.} \cite{DeSimone:2008if}.) 
It is reasonable to imagine that a particular brain configuration is highly specified, down to the level of
certain atomic or molecular states; nevertheless, real brains are wet, warm systems, and the state of
every single atom is certainly not crucial.  
A brain is generally not a ``bounce'' state; it will feature macroscopic motions that are not invariant 
under time-reversal.  Therefore, the most likely trajectories that form the brain will not be precisely the
time-reverse of the trajectories by which it decays.  However, we expect the crude features to be similar;
in particular, a series of unlikely tunneling and baryon-production events will gradually assemble into
a cold, dead brain, which will then ``come to life.''

It is interesting to speculate about the brain's psychological arrow of time in the last few seconds before
the configuration $A_0$ is reached.  Does the brain experience a reversed arrow of time, with memories of
the future?  In real brains, quantum processes occur and decohere on a very short timescale, generating
entropy along the way \cite{Tegmark:1999yf}.  It is at least conceivable that the brain's psychological arrow
remains synchronized with the thermodynamic arrow of its environment.  However, the macroscopic 
processes that characterize any particular brain state may be sufficiently robust that the psychological arrow
(to the extent that it is well-defined) remains consistent, and the pre-completion brain ``observes'' an 
environment of decreasing entropy.  In this context we will note that the question of whether the
``psychological'' arrow of time of an abstract computing machine must align with the thermodynamic arrow
is a controversial one \cite{e7040221,springerlink:10.1007/s10701-009-9386-6}.

We therefore have a scenario that looks like: (equilibrium) $\rar$ (eons of painstaking brain-assembly) $\rar$ (entropy-destroying brain without awareness) $\rar$ (possible phase of entropy destruction with awareness) $\rar$ (specified brain configuration $A_0$) $\rar$ (brief period marveling at the beauty of a pot of petunias and pondering the meaning of life \cite{adams}) $\rar$ (brain death, disintegration and equilibrium).

\section{Discussion}
\label{sec-discussion}

Observing a downward fluctuation in the entropy of a macroscopic isolated system on human timescales would appear to be a miraculous occurrence. However, where systems are small, and/or where timescales are ultra-long, the understanding of such fluctuations can be very important for a proper description of physical processes. In this paper, we have focused on the latter case, examining fluctuations that occur in gravitational systems that in some sense have a long-lived or eternal equilibrium state. The two most important examples are thermal AdS and dS space. We have clarified the details of a number of processes that occur in thermal AdS and dS, including: thermal, CDL, Lee-Weinberg, and HM vacuum transitions, stochastic eternal inflation, black/white hole nucleation, the ``Island Universe" scenario, and the fluctuation of ``Boltzmann brains."

Using elementary arguments in statistical mechanics, we have provided a formal prescription for determining the PDF over the histories of deviations from equilibrium: choose a non-equilibrium state $A_0$, take its CPT conjugate $\bar A_0$, and let it evolve back to equilibrium. The probability distribution over histories leading from equilibrium to $A_0$ is just the CPT conjugate of the probability distribution over histories leading from $\bar A_0$ to equilibrium. Our prescription applies under the assumptions that (a) we can define a system with a fixed set of states and their evolution (which requires additional assumptions when gravity is included), (b) this evolution is unitary, and in particular CPT or some other involution provides a conjugated state (and trajectory) for each state (and trajectory) in the system, and (c) that when a macrostate is specified at some time $t$, we assume a flat probability distribution at time $t$ over microstates consistent with that macrostate.

Our prescription should apply straightforwardly to terrestrial microscopic systems, but we have not pursued such applications; rather, we have focused on ultra-long timescales in which gravity and cosmology are important.   In doing so, we have assumed that the system is defined by the degrees of freedom necessary to describe the causal diamond of a single observer.  Our prescription indicates that many of the fluctuations typically discussed in the context of eternal inflation are not sudden or ill-defined, but rather require painstaking assembly. For example, to form a false vacuum bubble, it is necessary to re-play an entire cosmology in reverse.  This connects with the deep cosmological question of why our observable universe appears to have evolved from a much, much lower-entropy precursor.  For this to make sense, it seems that one of our assumptions -- that of a fixed set of states, or attainment of equilibrium, or democracy of microstates, or unitarity -- should be violated.  

This hinges, in turn, on exactly what we mean by the ``system", which becomes particularly subtle when considering spacetimes with a cosmological constant, where there are many regions which remain causally disconnected for all time. One can either interpret our prescription as making statements about the observable portion of some larger fluctuation, or, if the causal diamond is a complete description of all the physics, as making statements about fluctuations in a more fundamental sense. This local viewpoint is particularly helpful when trying to reconcile fluctuations on super-horizon scales with causality. However, there are a number of confusing aspects to confront. Perhaps most importantly, the definition of the causal diamond depends on the detailed history of a particular observer. While small fluctuations will not change the spacetime structure by much (in many cases, we have applied our prescription to the study of fluctuations in a fixed background), large ones can. This may not be particularly problematic in AdS, where boundary conditions are fixed. But in dS, especially in the case where there are vacuum transitions, it is unclear which observer to choose. One possible resolution is that different observers simply have access to different subsets of the states describing the whole system. This is the viewpoint advocated in~\cite{Banks:2005ru}. 

The wealth of insights obtained about semi-classical and quantum gravity from thermodynamic arguments highlights the importance of studying non-equilibrium phenomena. As we have argued above, equilibrium gravitational systems such as thermal AdS and dS space provide examples where such phenomena are important. This paper is a first step towards better understanding the full potential of this program.

\acknowledgements
We thank T.~Banks, A.~Brown and C.~Shalizi for useful comments. Supported in part by the U.S. Department of Energy and the Gordon and Betty Moore Fooundation, PHY-0757912,  and a ``Foundational Questions in Physics and Cosmology" grant from the John Templeton Foundation.
Research at Perimeter Institute is supported by the Government of Canada through Industry Canada and by the Province of Ontario through the Ministry of Research and Innovation. 

\bibliography{trevfluctb.bib}

\end{document}